\documentclass[longauth]{aa}

\usepackage{graphicx}
\usepackage{txfonts}
\usepackage{booktabs}
\usepackage[flushleft]{threeparttable}
\usepackage[table]{xcolor}
\usepackage{amsmath}
\usepackage{array}
\usepackage{float}
\definecolor{cyan}{rgb}{0.88,1.0,1.0}
\definecolor{yellow}{rgb}{1.0,1.0,0.0}
\definecolor{mypink}{rgb}{0.858, 0.188, 0.478}
\usepackage{multirow}
\usepackage{multicol}
\usepackage{xcolor}
\usepackage{blindtext}
\usepackage{subcaption}
\usepackage{longtable}
\usepackage{subcaption}
\newcommand{\teff}{$T_{\rm{eff}}$\,}
\newcommand{\logg}{$\log g$\,}
\newcommand{\kms}{km\,s$^{-1}$\,}
\newcommand{\vsini}{$v\,{\rm sin}\, i$}
\newcommand{\angstrom}{\mbox{\normalfont\AA}}

\usepackage{natbib,twoopt}
\usepackage[breaklinks=true]{hyperref} 
\bibpunct{(}{)}{;}{a}{}{,} 

\hypersetup{draft}
\begin{document}

   \title{The GAPS Programme at TNG \thanks{Based on observations made with the Italian Telescopio Nazionale
Galileo (TNG) operated on the island of La Palma by the Fundación Galileo Galilei of the INAF (Istituto Nazionale di Astrofisica) at the Spanish Observatorio del Roque de los Muchachos (ORM) of the IAC.}}

   \subtitle{XXV. Stellar atmospheric parameters and chemical composition through GIARPS optical and near-infrared spectra }

   \author{M.~Baratella\inst{\ref{unipd}}
          \and
          V.~D'Orazi\inst{\ref{oapd},\ref{monash}}
          \and
          K.~Biazzo\inst{\ref{oaro}}
          \and
          S.~Desidera\inst{\ref{oapd}}
          \and 
          R.~Gratton\inst{\ref{oapd}}
          \and
          S.~Benatti\inst{\ref{oapa}}
          \and
          A.~Bignamini\inst{\ref{trieste}}
          \and
          I.~Carleo\inst{\ref{carleo}}
          \and
          M.~Cecconi \inst{\ref{tng}}
              \and
          R.~Claudi\inst{\ref{oapd}}
          \and
          R.~Cosentino \inst{\ref{tng}}
          \and
          A.~Ghedina \inst{\ref{tng}}
          \and
          A.~Harutyunyan        \inst{\ref{tng}}
          \and
          A.F.~Lanza\inst{\ref{catania}}
          \and
          L.~Malavolta\inst{\ref{unipd}}
          \and
          J.~Maldonado\inst{\ref{oapa}}
          \and
          M.~Mallonn\inst{\ref{mallonn}}
          \and
          S.~Messina\inst{\ref{catania}}
          \and
          G.~Micela \inst{\ref{oapa}}
              \and
              E.~Molinari \inst{\ref{Cagliari}}
          \and 
          E.~Poretti\inst{\ref{Brera},\ref{tng}}
          \and
          G.~Scandariato \inst{\ref{catania}}
          \and
          A.~Sozzetti\inst{\ref{oato}}
          }

   \institute{Dipartimento di Fisica e Astronomia {\it Galileo Galilei}, Vicolo Osservatorio 3, I-35122, Padova, Italy \label{unipd}\\
\email{martina.baratella.1@phd.unipd.it}
    \and INAF -- Osservatorio Astronomico di Padova, vicolo dell'Osservatorio 5, 35122, Padova, Italy \label{oapd}
  \and School of Physics and Astronomy, Monash University, Clayton, VIC 3800, Melbourne, Australia\label{monash}
  \and INAF -- Osservatorio Astronomico di Roma, via Frascati 33, I-00040, Monte Porzio Catone (RM), Italy \label{oaro}
  \and INAF -- Osservatorio Astronomico di Palermo, Piazza del Parlamento 1, I-90134, Palermo, Italy \label{oapa}
  \and INAF -- Osservatorio Astrofisico di Torino, Via Osservatorio 20, I-10025 Pino Torinese, Italy \label{oato}
  \and INAF -- Osservatorio Astronomico di Trieste, via Tiepolo 11, 34143 Trieste, Italy \label{trieste}
  \and Astronomy Department, 96 Foss Hill Drive, Van Vleck Observatory 101, Wesleyan University, Middletown, CT 06459, US \label{carleo}
  \and INAF -- Osservatorio Astrofisico di Catania, Via S. Sofia 78, 95123,Catania, Italy \label{catania}
  \and INAF -- Osservatorio Astronomico di Brera, Via E. Bianchi 46, 23807 Merate (LC), Italy \label{Brera}
  \and Fundaci\'on G. Galilei - INAF (Telescopio Nazionale Galileo), Rambla J. A. Fern\'andez P\'erez 7, E-38712 Bre\~na Baja (La Palma), Spain \label{tng}
  \and Leibniz-Institut f\"{u}r Astrophysik Potsdam (AIP), An der Sternwarte
16, D-14482 Potsdam, Germany \label{mallonn}
  \and INAF – Osservatorio di Cagliari, via della Scienza 5, I-09047 Selargius, CA, Italy \label{Cagliari}
  \\}

   \date{}
   
      \date{Received ??; accepted ??}

 
  \abstract
   {The detailed chemical composition of stars is important in many astrophysical fields, among which is the characterisation of exoplanetary systems. Previous studies seem to indicate an anomalous chemical pattern of the youngest stellar population in the solar vicinity that has sub-solar metal content. This can influence various observational relations linking the properties of exoplanets to the characteristics of the host stars, for example the giant planet-metallicity relation.   }
   {In this framework, we aim to expand our knowledge of the chemical composition of  intermediate-age stars and understand whether these peculiarities are real or related to spectroscopic analysis techniques. }
   {We analysed high-resolution optical and near-infrared spectra of intermediate-age stars ($< 700$\,Myr) that have been observed simultaneously with HARPS-N and GIANO-B spectrographs in GIARPS mode. To overcome issues related to the young ages of the stars, we applied a new spectroscopic method that uses titanium lines to derive the atmospheric parameters, in particular surface gravities and microturbulence velocity parameter. We derived abundances of C\,{\sc i}, Na\,{\sc i}, Mg\,{\sc i}, Al\,{\sc i}, Si\,{\sc i}, Ca\,{\sc i}, Ti\,{\sc i}, Ti\,{\sc ii}, Cr\,{\sc i}, Cr\,{\sc ii},  Fe\,{\sc i}, Fe\,{\sc ii}, Ni\,{\sc i,} and Zn\,{\sc i}. }
   {The lack of systematic trends between elemental abundances and effective temperatures validates our methods. However, we observed that the coolest stars in the sample, where \teff$<5400$\,K, display higher abundances for the ionised species, in particular Cr\,{\sc ii}, and for high-excitation potential C\,{\sc i} lines.  }
  {We found a positive correlation between the higher abundances measured of C\,{\sc i} and Cr\,{\sc ii} and the activity index  $\log $R$^{\prime}_{\rm{HK}}$. Instead, we found no correlations between the C abundances obtained from CH molecular band at 4300\AA\, and both effective temperatures and activity. Thus, we suggest that these are better estimates for C abundances in young and cool stars. Finally, we found an indication of an increasing abundance ratio [X/H] with the condensation temperature for HD\,167389, indicating possible episodes of planet engulfment.}


   \maketitle
%

\section{Introduction}\label{introduction}

The precise determination of the atmospheric parameters and chemical composition of stars plays a crucial role in many astrophysical fields. In particular, this is of primary importance for exoplanetary studies to fully understand the main observational correlations between the properties of exoplanets and the characteristics of their host star; these correlations include the giant planet-metallicity relation and the trends observed between the condensation temperature (Tc) and abundances ratios [X/Fe] (see e.g. \citealt{2015nissen,2016brewer,2019adibekyan} and references therein). 

High-resolution stellar spectroscopy is one of the most powerful tools to fully characterise a star. This technique allows us to determine with great precision the physical properties of stars, for example effective temperatures (\teff), and the chemical abundances of various atomic and molecular species (see \citealt{2019jofre} for a complete review). However, in some cases the spectroscopic analysis of stars is not trivial and the results can be affected by multiple issues.
In particular, young stars ($<200$\,Myr) have higher activity levels, which can alter the structure of the atmosphere, complicate the derivation of the stellar parameters and consequently of the metallicity, and can hamper the detectability of planets (e.g. \citealt{2018carleo,2020carleo}).

\begin{table*}[!ht]
\small{
\caption{Some basic information of both standard stars and members in young associations analysed in this work. }
\label{tabinfo}
\centering
\renewcommand\arraystretch{1.0}
\setlength\tabcolsep{3.5pt}
\begin{threeparttable}
\begin{tabular}{lccccccccccccccccc}
\toprule
SIMBAD ID  & RA & Dec & SpT & V  & J & H & K  & d$^a$ & Age  & \vsini & $\log $R$^{\prime}_{\rm{HK}}$\\
&\footnotesize{(J2000)} & \footnotesize{(J2000)} & & \footnotesize{(mag)}&\footnotesize{(mag)}&\footnotesize{(mag)}& \footnotesize{(mag)} & \footnotesize{(pc)} &\footnotesize{(Gyr)}  & \footnotesize{(km\,s$^{-1}$)}\\
\midrule
HD\,3765 & 00 40 49.27 & +40 11 13.82 & K2 & 7.344 & 5.694 & 5.272 & 5.164 & 17.94$\pm$0.03 & 4.94$\pm$6.56  & 2.6$^e$ & $-$4.94$\pm$0.01\\
HD\,159222 &  17 32 00.99 & +34 16 16.13 & G1 & 6.595 & 5.342 & 5.076 & 4.998 &   24.22$\pm$0.01 & 3.24$\pm$1.48 &  3.01$^f$ & $-$4.88$\pm$0.01  \\
\\
\hline
& & & & &\bf{Coma Berenices}\\
TYC\,1991-1235-1 &       12 28 56.43 & +26 32 57.39 & K5 &  10.971  & 9.208 & 8.768 & 8.661 & 84.14$\pm$0.33 & 0.56$\pm$0.09$^b$ &  3.5$\pm$1.2$^g$ & $-$4.41$\pm$0.05 \\
HIP\,61205 &   12 32 31.07 & +35 19 52.31 & G0 &  9.635 & 8.407 & 8.132 & 8.086  & 83.41$\pm$0.32 & 0.56$\pm$0.09 &  6.4$\pm$0.9$^i$ & $-$4.43$\pm$0.02 \\
TYC\,1989-0049-1 &   12 21 15.62 & +26 09 14.05 & K3 & 11.483 & 9.614 & 9.087 & 8.972 & 84.71$\pm$0.29 & 0.56$\pm$0.09 &  1.4$\pm$2.8$^g$ & $-$4.17$\pm$0.02\\
TYC\,1989-147-1 &        12 24 05.73 & +26 07 42.92 & K0 & 10.461 & 9.081 & 8.762 & 8.611 & 88.67$\pm$0.31 & 0.56$\pm$0.09 &  5.0$\pm$0.9$^g$& $-$4.55$\pm$0.01\\
\\
\hline
& & & & &\bf{Ursa Major}\\
HD\,167389 &  18 13 07.23 & +41 28 31.31 & F8 & 7.453 & 6.224 & 5.968 & 5.918 & 34.72$\pm$0.03 & 0.5$\pm$0.1$^c$ &  3.5$\pm$0.5$^l$ & $-$4.78$\pm$0.02\\
HD\,59747  &07 33 00.58 & +37 01 47.45 & G5 & 7.797 & 6.090 & 5.662 & 5.589 & 20.68$\pm0.02$ & 0.5$\pm$0.1 &   2.6$\pm$0.5$^h$ & $-$4.37$\pm$0.02 \\
\hline
& & & & &\bf{Hercules Lyra}\\
HD\,70573 &  08 22 49.95 & +01 51 33.55 & G1&  8.711 & 7.558 & 7.276 & 7.191 & 59.28$\pm$0.16 & 0.25$\pm$0.05$^d$ &  13.5$\pm$0.5$^m$ & $-$4.31$\pm$0.02\\
\hline
\end{tabular}
\tablebib{
a) \cite{2018bailer}; b) \cite{2014silaj}; c) \cite{2001montes}; d) \cite{2013eisenbeiss}; e) \cite{2017luck}; f) \cite{2010martinez}; g) \cite{2009mermilliod}; h) \cite{2014marsden}; i) \cite{2008mermilliod}; l) \cite{2005valenti}; m) \cite{2010gonzalez}   }
\begin{tablenotes}
\small
\item 
\item 
\end{tablenotes}
\end{threeparttable}
}
\end{table*}

The presence of active chromospheres or intense photospheric magnetic fields \citep{2016folsom} may alter the spectral line formation. Recently, \cite{2020baratella} show that the apparent sub-solar metallicity observed for the young stars in the solar neighbourhood may be related to an over-estimation of the microturbulence velocity ($\xi$) parameter. This is a free fictitious parameter representing small-scale motions of matter in the photospheric layers of the star and it is introduced in 1D spectroscopic analysis to account for the difference between the observed and predicted equivalent widths (EWs), when models account only for thermal and damping broadening.  Weaker lines are less affected by this parameter, which is calculated by forcing lines, usually of iron (Fe), to give the same abundance. However, higher values of $\xi$ lead to systematic under-estimation of the elemental abundances, so that the stars belonging to young associations and open clusters (OCs; $ <$ 200\,Myr) could appear more metal-poor than what it is predicted by Galactic chemical evolution models \citep{2011dorazi,2011biazzoA,Spina17}. For example, \cite{2006james} analysed young stars observed in three star-forming regions ($\tau \sim$ 10 Myr) and reported extremely high $\xi$ values for pre-main-sequence stars of up to 2.5 \kms, which is expected for more evolved stars. Similar results have been also reported by \cite{2008santos}. \cite{2009viana} analysed stars in 11 young associations (ages less than 100\,Myr) and again found $\xi$ values up to 2.6\,\kms. Moreover, they reported a small trend of increasing $\xi$ at decreasing \teff.
Several other authors confirmed the anomalous values of $\xi$ found for young stars, but they reported close-to-solar values of the metallicity. For instance, \cite{2009dorOrion} and \cite{2011biazzoA,2011biazzoB} reported values up to $\sim$2\,\kms for stars belonging to the Orion complex. However, for some stars they also reported values of metallicity varying from $-0.15$ to 0.01\,dex, despite the large scatter in the $\xi$ parameter measurements (e.g. in \citealt{2011biazzoB} a star in $\lambda$ Orionis has $\xi$=2.1\,\kms, but [Fe/H]=0.00\,dex).

Recently, \cite{2019galarza} show that the observed EWs of iron lines vary with the activity phase of the young solar analogue HD\,59967 (age $\sim$ 400\,Myr). In particular, they find that the line strength increases when the star is more active, producing variations of $\xi$ and iron abundance along with the stellar cycle. Moreover, these authors demonstrate that such variations of EWs depend on the optical depth of line formation and, marginally, on the Landè $g_{\rm{L}}$  factor, which measures the sensitivity of a spectral line to magnetic fields. \cite{2020spina} conducted the same study as \cite{2019galarza} to a sample of 211 solar-analogue stars observed with High Accuracy Radial velocity Planet Searcher (HARPS) and find similar results.  \cite{2019galarza}, \cite{2020baratella}, and \cite{2020spina} demonstrate how iron lines forming in the upper layers of the photosphere of young stars can yield higher abundances due to the possible influence of the more intense chromospheric or photospheric magnetic fields.

Higher levels of stellar activity can also affect the abundances of some elements when they are derived using high-excitation potential lines (see \citealt{2015schuler} and references therein). It has been observed that young ($ <$ 200\,Myr) and cool (\teff$\lesssim$ 5400\,K) stars display anomalous abundances of oxygen O\,{\sc i} triplet ($\chi = 9.15$\,eV) and sulfur (S\,{\sc i} line at 6053\AA\, with $\chi = 7.87$\,eV) \citep{2004schuler,2013teske,2013ramirez}. In particular, the abundances increase at decreasing \teff, reaching values of 0.8-1.0\,dex over solar for the coolest stars (\teff $\sim$4700\,K). Similarly, for the same kind of stars and in the same \teff regime, differences between the neutral and ionised species of the same element of the order of +0.8\,dex have been observed for Fe and Ti \citep{2006schuler,2009dorazi}. Such differences can produce unreliable results, in particular for the derivation of \logg, which should be decreased in order to satisfy the ionisation equilibrium. These effects  may be caused either by non-local thermodynamic equilibrium (NLTE) departures, for which the high-energy levels are not correctly modelled, or by the presence of unidentified blends (see also \citealt{2019tsantaki}), or a combination of both. \cite{2017aleo} argue that the large differences between Fe\,{\sc i} and Fe\,{\sc i} may be related to blending of Fe\,{\sc ii} lines that become more severe at decreasing \teff. These results were corroborated by \cite{2020takeda}, who also concluded that the O\,{\sc i} overabundance obtained from the oxygen triplet by \cite{2006schuler} might be due to the different \teff scale and to over-estimation of the strength of the lines in coolest stars.
Even though we are aware of the issues related to the spectroscopic analysis of young and cool stars and we are starting to shed light on the topic, we still lack a definitive solution. However, we can overcome these problems with strategic choices of the line list to use in the analysis, for example with a refined selection based on the EWs \citep{2020spina} or using new approaches \citep{2020baratella}.

Thanks to the advent of large spectroscopic surveys, such as \textit{Gaia}-ESO Survey \citep{2012gilmore} or GALactic Archaeology with HERMES (GALAH; \citealt{2015galah}), the number of stellar spectra has increased enormously. Along with the increasing number of available spectra, the need arose to assess the precision and accuracy of spectroscopic analysis techniques \citep{2019jofre}. Until recently, the study of stellar spectra mainly involved the analysis of data in the optical band, covering the wavelength range from $\sim$4000 to $\sim$7000\,\AA. However, the advent of high-resolution near-infrared (NIR) spectroscopy allowed us to extend the analysis of stellar spectra at longer wavelengths as well, and to test the validity of optical and NIR analysis techniques \citep{2020marfil}. This is particularly important in the study of young and intermediate-age stars, for which stellar activity and other effects can alter the derivation of atmospheric parameters and, specifically, the chemical composition.

For five years, the Global Architecture of Planetary Systems (GAPS) project \citep{2013covino,2016poretti} searched for planets through a radial velocity (RV) technique with High Accuracy Radial velocity Planet Searcher for the Northern emisphere (HARPS-N; \citealt{2014cosentino}) at Telescopio Nazionale Galileo (TNG, Roque de los Muchachos, La Palma) around different types of stars, including the characterisation of selected planet-host stars. Recently, a new phase of the project started with the aim of exploiting the full capabilities of the GIARPS mode \citep{2017claudi}. This means that we can study
and fully characterise planetary systems by analysing GIAno-B \citep{2006oliva} and haRPS stellar spectra acquired simultaneously. In this context, the GAPS Young Objects (GAPS-YO) project \citep{2020carleo} aims to monitor and study young ($<100$\,Myr) and intermediate-age ($<700$\,Myr) stars to search for and characterise hot and warm planets down to sub-Neptune mass in formation or at an early stage of their evolution.

In this first paper of a series, we present the results of spectral characterisation in terms of astrophysical parameters and elemental abundances of stars observed within the GAPS-YO project. Our analysis includes the Sun, two RV standard stars (HD\,3765 and HD\,15922), and seven more stars members of intermediate-age stellar clusters and moving groups. In Sec.2 we present the data we analysed and we report some information on the selected stars. Our analysis is separated between the optical and NIR spectral ranges. In particular, in Sec. 3 we describe the new method applied to derive atmospheric parameters and elemental abundances from optical spectra. These parameters were used to derive abundances of neutral C, Na, Mg, Al, Si, Ca, Ti, Fe, and Ni from NIR spectral lines (Sec. 4). In Sec. 5 we present the resulting chemical abundances of various atomic species and discuss the scientific implications. In Sec. 6 we present our conclusions.

\begin{table}[!]
\caption{Solar abundances derived in the present work from the analysis of HARPS-N and GIANO-B spectra. We also report the values from \cite{asplund} (A09) for comparison.}
\setlength\tabcolsep{4pt}
\centering
\begin{tabular}{lcccc}
\toprule
Species & HARPS-N & GIANO-B & A09\\
\midrule
C\,{\sc i} & 8.45$\pm$0.04\tiny{(NLTE)} & 8.38$\pm$0.10  & 8.43$\pm$0.05 \\
Na\,{\sc i} & 6.21$\pm$0.04\tiny{(NLTE)} & 6.24$\pm$0.04 & 6.24$\pm$0.04 \\
Mg\,{\sc i} & 7.63$\pm$0.04 & 7.59$\pm$0.01 & 7.60$\pm$0.04 \\
Al\,{\sc i} & 6.49$\pm$0.03 & 6.45$\pm$0.03 & 6.45$\pm$0.03\\
Si\,{\sc i}& 7.54$\pm$0.02 & 7.52$\pm$0.01 & 7.51$\pm$0.03 \\
Ca\,{\sc i} & 6.35$\pm$0.05 & 6.36$\pm$0.01 & 6.34$\pm$0.04 \\
Ti\,{\sc i} & 4.97$\pm$0.02 & 4.98$\pm$0.01 & 4.95$\pm$0.05 \\
Ti\,{\sc ii} & 4.98$\pm$0.04 & -&\\
Cr\,{\sc i} & 5.65$\pm$0.04 & - & 5.64$\pm$0.04\\
Cr\,{\sc ii} & 5.66$\pm$0.05 &- &\\
Fe\,{\sc i} & 7.49$\pm$0.03 & 7.51$\pm$0.01 & 7.50$\pm$0.04 \\
Fe\,{\sc ii} & 7.48$\pm$0.04 & - \\
Ni\,{\sc i} & 6.24$\pm$0.04 & 6.22$\pm$0.02 & 6.22$\pm$0.04 \\
Zn\,{\sc i} & 4.55$\pm$0.01 & -& 4.56$\pm$0.05 \\
\hline
\end{tabular}
\label{solar}
\end{table}

\section{Sample selection and spectroscopic data}

In this work, we analysed high-resolution spectra of seven young and intermediate-age stars observed in the GAPS-YO project. We selected spectra with high signal-to-noise ratios (S/N),  low rotational velocities ( \vsini$<15$\kms) to avoid line blending, and spectral type F-G-K. We excluded from the analysis stars with spectral types later than K to avoid problems with the molecular bands. The selected targets are as follows:  TYC\,1991-1235-1, HIP\,61205, TYC\,1989-0049-1, and TYC\,1989-147-1, which belong to the Coma Berenices OC \citep{2008mermilliod}, with an age of $\sim$600\,Myr; HD\,167389 and HD\,59747, which are part of the Ursa Major moving group \citep{2001montes}, with an age of $\sim$500\,Myr; and  HD\,70573 of Hercules Lyra moving group \citep{2006lopez}, with an age of $\sim$200\,Myr. We also analysed, for validation, the spectra of two old stars observed as RV standard stars, HD\,3765 and HD\,159222. We reported some information on the selected targets in Table \ref{tabinfo}.

The spectra were acquired with HARPS-N and GIANO-B spectrographs placed at the 3.6 m INAF-Telescopio Nazionale Galileo (TNG) in La Palma. The HARPS-N spectrograph is the northern counterpart of HARPS at the La Silla Observatory (Chile), mounted at the Nasmyth-B focus of the TNG. With a resolving power R$\sim$115000 and large wavelength coverage in the optical range (0.38-0.69 $\mu$m), it allows us to obtain very precise (less than 1 m\,s$^{-1}$) RV measurements, thanks to an accurate control system that minimises pressure and temperature variations  and prevents spectral drifts due to environmental conditions. The GIANO-B spectrograph is a high-resolution (R$\sim$45000-50000) NIR spectrograph covering the wavelength range from 0.95 $\mu$m to 2.45 $\mu$m that is placed at Nasmyth-B focus of the TNG. 
The configuration of the two spectrographs allows us to observe the stars simultaneously in the optical and NIR wavelengths in the GIARPS mode. We analysed GIARPS spectra of the Sun, HD\,3765, HD\,159222, TYC\,1991-1235-1, HIP\,61205, and HD\,167389. We also analysed the optical spectra of the remaining four additional targets.

HARPS-N data are reduced with the standard Data Reduction Software (DRS). Since the spectra were collected by the GAPS-YO collaboration to obtain time series for RV monitoring, the available HARPS-N data for each target were then combined to obtain a co-added spectrum with S/N $>$ 100 \citep{2016malavolta}.
The NIR data reduction was performed with the pipeline GOFIO \citep{2018gofio,2018harutyunyan}, while the telluric correction was performed following the method described in \cite{2016carleo}. We verified that co-adding HARPS-N spectra did not introduce any systematic errors, thanks to the high stability of the instrument over several months. This was not the case for GIANO-B spectra, for which we decided to consider the highest S/N observation (S/N>70) for each star. Since we performed a differential analysis with respect to the Sun, we derived our solar abundance scale by analysing the HARPS-N and GIANO-B spectra of Ganymede, which have a S/N = 145 at 607\,nm and S/N=180 at 1500\,nm, respectively.

\section{Optical analysis}

For the analysis of the HARPS-N optical spectra, we employed the same approach as in \cite{2020baratella}, which exploits the use of Ti lines to derive the atmospheric parameters. On average, Ti lines form deeper in the photosphere than Fe lines, so they are less affected by the chromosphere, which is more active in young stars. In this way, we can overcome the issues affecting the analysis of young stars, which have been already presented in Sec.\ref{introduction}. Briefly, the new spectroscopic method is based on the use of Ti and Fe lines to derive \teff by imposing the excitation equilibrium and the use of Ti lines only to derive \logg and $\xi$ by imposing the ionisation equilibrium and by removing the trend between the single line abundances and the reduced equivalent width (REW\footnote{REW=$\log$(EW/$\lambda$)}), respectively.

For the analysis, we used the local thermodynamic equilibrium (LTE) code MOOG\footnote{https://www.as.utexas.edu/~chris/moog.html} (version 2017, \cite{sneden73}; \cite{2011sob}). We estimated the abundances of C\,{\sc i}, Na\,{\sc i}, Mg\,{\sc i}, Al\,{\sc i}, Si\,{\sc i}, Ca\,{\sc i}, Ti\,{\sc i}, Ti\,{\sc ii}, Cr\,{\sc i}, Cr\,{\sc ii}, Fe\,{\sc i}, Fe\,{\sc ii}, Ni\,{\sc i,} and Zn\,{\sc i} using the EW method by running the \textit{abfind} driver. We adopted the same line list used in \cite{2020dorazi}  that includes 86 Fe\,{\sc i} lines, 17 Fe\,{\sc ii} lines, 57 Ti\,{\sc i} lines, 22 Ti\,{\sc ii} lines, and 42 more lines of different atomic species;  Table A.1 provides
a complete line list with the atomic data. We added two C\,{\sc i} lines to the original line list, taking into account the atomic data from \cite{2019amarsi}, specifically lines 5380.34 and 6587.61 \angstrom.  We used the Barklem prescriptions for damping values (see \citealt{2000barklem} and references therein).

We measured EWs for all lines via the software ARESv2 \citep{sousa}\footnote{http://www.astro.up.pt/~sousasag/ares/}, which calculates EWs through a Gaussian fitting of the line. We discarded the lines with fitting errors larger than 10\% and those lines with EWs$>$120\,m\AA\,. In this way, we removed strong lines for which the Gaussian approximation is not adequate. We used 1D model atmospheres linearly interpolated from the ATLAS9 grid of Castelli \& Kurucz (\citeyear{2003cast}), with new opacities (ODFNEW). We estimated the input values of the \teff and of surface gravities (\logg) as in \cite{2020baratella}. The \teff estimates were obtained via 2MASS photometry \citep{2003cutri} in the calibrated relation by \cite{casagrande} that is valid for $(J-K)$ de-reddened colours. The initial values of the surface gravities (trigonometric gravities, \logg$_{\rm{trig}}$) were estimated using the classical equation, based on Gaia DR2 distances as calculated by \cite{2018bailer} (see Table \ref{tabinfo}). Instead, the initial values of $\xi$ were derived using the relation by \cite{ferreira}, calibrated for dwarf stars, that is

\begin{equation}
\begin{split}
\xi(km\,s^{-1}) = & 0.998+3.16\times10^{-4}\,X-0.253\,Y\\
& -2.86\times10^{-4}\,X\,Y+0.165Y^2
\end{split}
,\end{equation}

where $X=T(J-K)-5500$\,(K) and $Y=$\logg $- 4.0 $ \,(dex).
In the calibrated relation used to derive the \teff, the input metallicity was assumed to be solar, which was later confirmed by the chemical abundances analysis. 

For the derivation of \teff we required that the slope of the trend between the Fe + Ti individual line abundances and $\chi$ is lower than its error. We adopted the same criteria for $\xi$, derived from the relation between Ti lines abundances and REWs. Instead, for the \logg we required that the difference between Ti\,{\sc i} and Ti\,{\sc ii} is lower than the quadratic sum of the errors on the abundances as calculated by MOOG. The uncertainties on \teff and $\xi$ were calculated by varying each quantity until the slopes of the relative trends are larger than their errors, while for \logg it was calculated by varying these parameters until the difference between neutral and ionised species is larger than the total error.  The uncertainties on the abundances include the internal errors due to EWs measurements ($\sigma_1$) and the contribution of the atmospheric parameters ($\sigma_2$), which is calculated by varying \teff, \logg, and $\xi$ one by one by their uncertainties, and calculating the difference with the new abundances. 

For the solar atmospheric parameters, we obtained \teff=5790$\pm$75\,K, \logg=4.40$\pm$0.05\,dex and $\xi$=0.93$\pm$0.05\,\kms. We reported the solar abundances of each element in Table\,\ref{solar}, where the uncertainties are the quadratic sum of the $\sigma_1$ and $\sigma_2$ contributions. As shown, our abundances are in very good agreement with the solar abundances of \cite{asplund}. The final values of atmospheric parameters and the derived abundances of neutral and ionised Fe and Ti for the stars in our sample are reported in Table \ref{param}.
We also calculated the abundance ratios [X/Fe] as [X/Fe]=[X/H]$_{\star}-$[Fe/H]$_{\star}$ (in particular, for the ionised species [X/Fe]$_{\rm{II}}$ = [X/H]$_{\rm{II}}-$[Fe/H]$_{\rm{II}}$). The final abundance ratios are reported in Table \ref{ratios}: for star HD\,70573, we could not derive the abundances of C and Al because of the relatively high \vsini\,\,. The analysis of HD\,3765 was not trivial. We derived \teff=5001$\pm$75\,K, but if the ionisation equilibrium was satisfied for Ti, this was not the case for Fe and Cr. In particular, we obtained a difference of +0.11 and +0.18\,dex between ionised and neutral Fe and Cr species, respectively. The same issue was raised by \cite{2007ramirez}, who obtained a difference between Fe\,{\sc i} and Fe\,{\sc ii} of +0.18\,dex. Moreover, we also obtained an anomalously large value of [C/H]=+0.36$\pm$0.05$\pm$0.09\,dex. Similar values of carbon abundances were also obtained for the other cool stars in our sample. This behaviour is discussed extensively in Section 5.

\begin{figure}[!]
\centering
\includegraphics[width=0.5\textwidth]{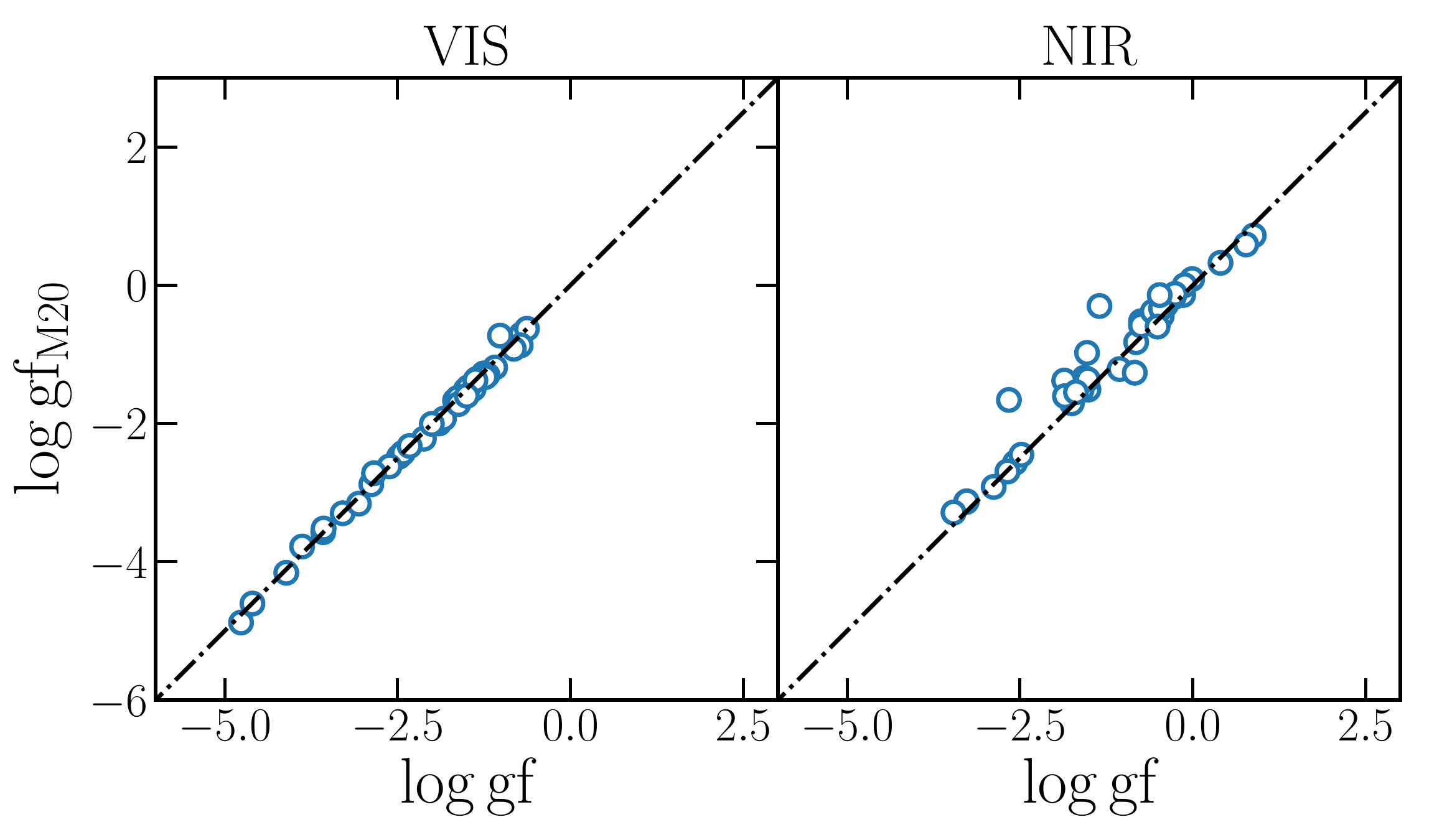}
\caption{Comparison of $\log gf$ between our line lists and those used in \cite{2020marfil} (M20).  }
\label{conf_loggf}
\end{figure}

\section{Near-infrared analysis}

The GIANO-B spectra were acquired for HD\,3765, HD\,159222, TYC\,1991-1235-1, HIP\,61205 and HD\,167389.  The NIR abundances were measured through spectral synthesis via the driver \textit{synth} in MOOG. We measured abundances of Na\,{\sc i}, Mg\,{\sc i}, Al\,{\sc i}, Si\,{\sc i}, Ca\,{\sc i}, Ti\,{\sc i}, Fe\,{\sc i,} and Ni\,{\sc i} using the same line list as \cite{2020dorazi}. Moreover, we added one C\,{\sc i} line to the original line list, the 16021.7\,\AA\, line, for which atomic data were taken from \cite{2015shetrone}.  Since in general the number of lines of the ionised atomic species is significantly lower in the NIR than in the optical part of the spectrum, we used the atmospheric parameters derived from the analysis in the optical part. To derive the abundances, we synthesised a region of 1000\,\AA\, and determined the best instrumental profile. Then we focussed on the line of interest and we derived the given abundance with 0.1\,dex steps to find the best-fit profile that minimises the sum of the squared residuals between the synthetic and the observed spectra. In Table \ref{solar} we reported the mean values obtained from the analysis of the solar spectrum.  As shown, the agreement with the optical values is extremely good, also validating the results of the atmospheric parameters we obtained from the optical analysis. The final abundances for the stars for which we analysed GIANO-B spectra are reported in Table \ref{nir}. For C, Na, Al, and Mg, we measured only one line, so the uncertainties $\sigma_1$ and $\sigma_2$ on the abundances in the Table account for the uncertainties on the fitting procedure and the sensitivity of [X/H] to changes in the atmospheric parameters, respectively. Instead, for the remaining elements for which we measured more than one line, we reported the mean values of the abundances, where $\sigma_1$ is the error on the mean and $\sigma_2$ related to the atmospheric parameters, respectively.

Recently, \cite{2020marfil} (hereafter M20) analysed CARMENES spectra of a sample of F-G-K stars (wavelength coverage between 5200-17100\,\AA\, and R=95000 and R=80000 in the optical and NIR channel, respectively) with the EWs method, using an extended line list that also comprises Fe\,{\sc i} and Fe\,{\sc ii} lines in the NIR part (216 and 1 lines, respectively) to assess the impact of the NIR lines on the derived stellar parameters. We compared the $\log gf$ values of our line list and those of the authors and we find that the values are nearly the same, as shown in Fig.\ref{conf_loggf}. The mean difference between our values and those in M20 is +0.02$\pm$0.08 for optical and $-$0.13$\pm$0.26 for NIR; thus we expect to obtain the same results as M20. We applied to our solar spectrum the same procedure and the same line list (the one that the M20 authors optimised for metal-rich dwarfs) as in M20. Since the wavelength coverage of CARMENES spectrograph is different than that of by HARPS-N and GIANO-B, we measured a total of 165 Fe\,{\sc i} and Fe\,{\sc i} lines adopted from M20 in the solar spectrum, in particular 125 lines in VIS and 40 lines in NIR. The analysis of optical+NIR spectra produced \teff=5790$\pm$50\,K, \logg=4.50$\pm$0.10\,dex, $\xi$=0.70$\pm$0.10\,\kms, with $\log$(Fe\,{\sc i})=7.53$\pm$0.01$\pm$0.04 and $\log$(Fe\,{\sc ii})=7.54$\pm$0.05$\pm$0.04. These values are very similar to what we obtained from the optical analysis alone and using our line list. The M20 authors, instead, find nearly the same \teff and \logg, but $\xi$=1.31$\pm$0.09\,\kms. We derived the atmospheric parameters applying the same criteria as the code STEPAR \citep{2019tabernero} and we find \teff=5750$\pm$75\,K, \logg=4.40$\pm$0.05\,dex and $\xi$=0.77$\pm0.15$\,\kms, confirming what we previously found. We believe that such a large discrepancy of the $\xi$ values is mainly due to the EW measurements of lines in the NIR. The NIR may pose a challenge when it comes to measuring EWs: for example, telluric lines in emission that remain after the correction, which that the placement of the continuum and a variable S/N ratio that is smaller at shorter wavelengths (Marfil, priv. comm.). Thus, it is possible that we measured different EWs than M20 and these differences may be responsible for the discrepancy in $\xi$ values.

\begin{figure*}[!t] 
\centering
\includegraphics[width=1.0\textwidth]{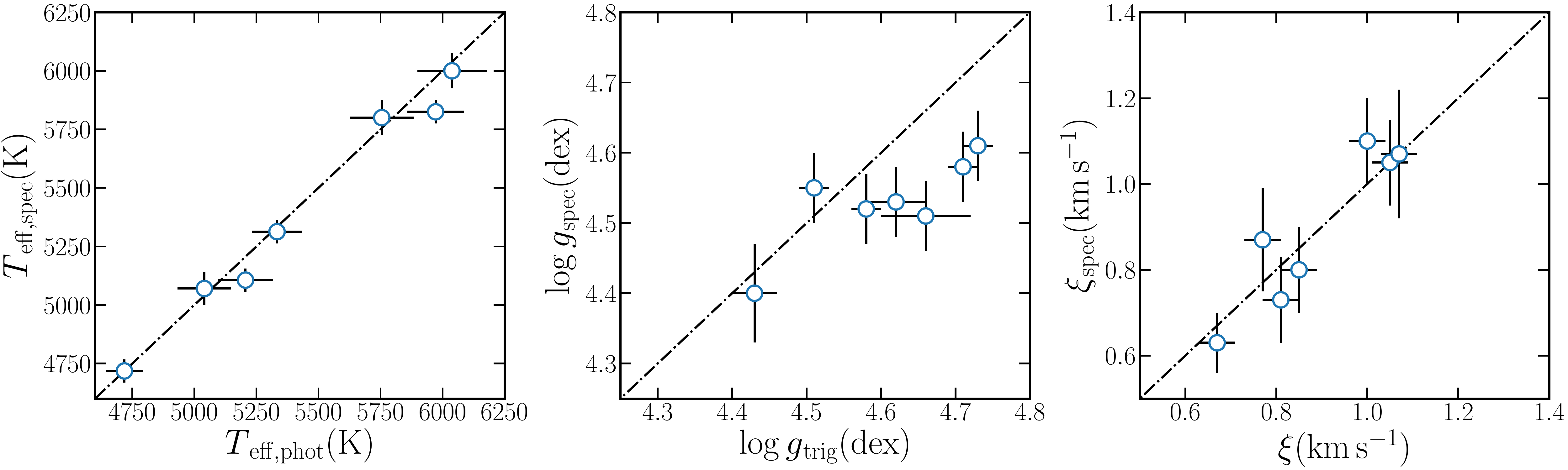}
\caption{Comparison between the input estimates of the atmospheric parameters and the derived spectroscopic values. The dash-dotted line represents the 1:1 relation. }
\label{conf_param}
\end{figure*}

\begin{figure}[!]
\centering
\includegraphics[width=0.40\textwidth]{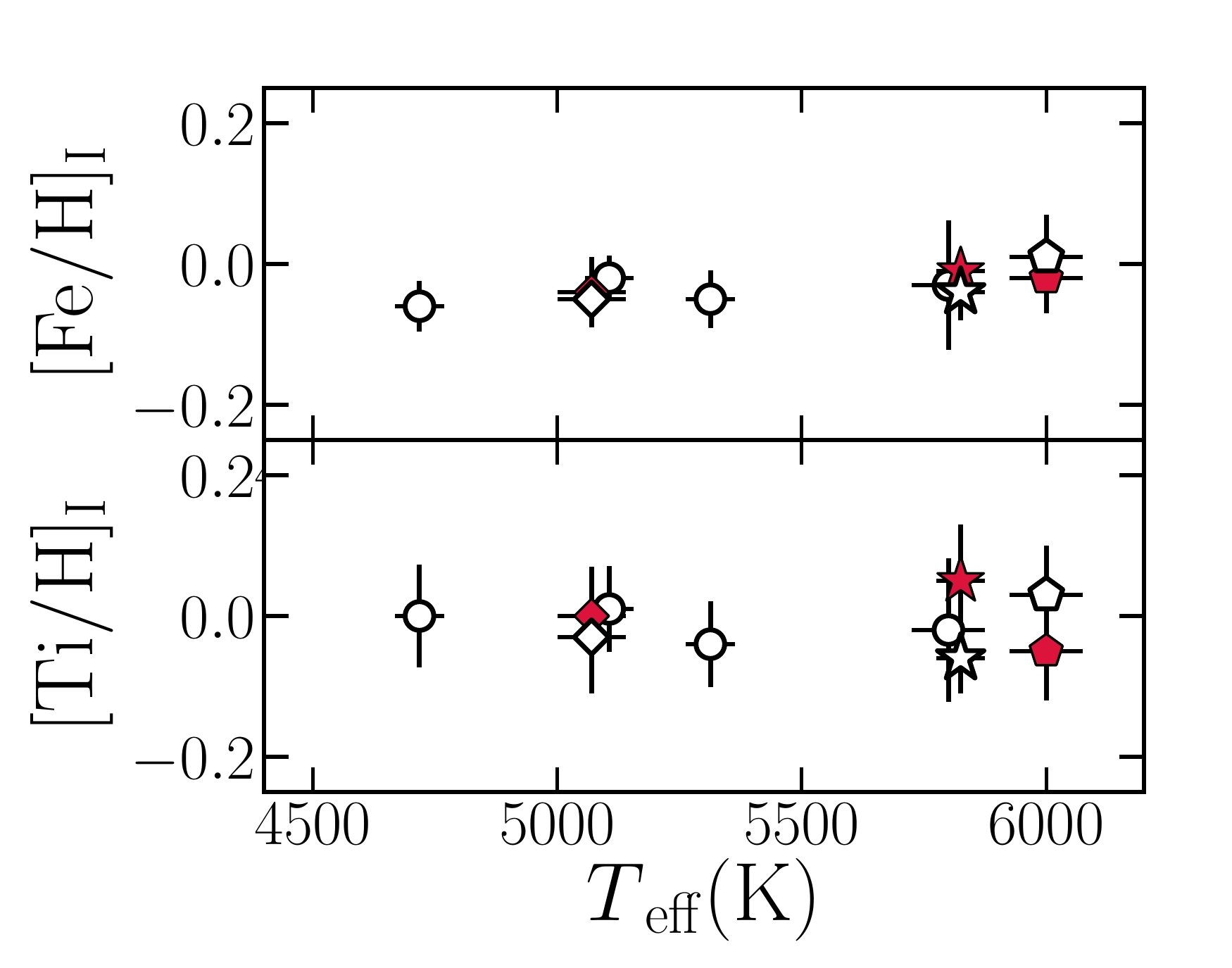}
\caption{Abundances of Fe\,{\sc i} and Ti\,{\sc i} as function of \teff. The open symbols represent the optical measurements, while the red symbols indicate the NIR measurements. The diamond indicates TYC\,1991-1235-1, the star symbol represents HIP\,61205, and the pentagon denotes HD\,167389, for which we have both optical and NIR measurements. The Pearson correlation coefficient of the trend in the top panel is $r$=0.67, which is not significant at $p<$0.1. For Ti the Pearson correlation coefficient is $r$=$-$0.03, which is not significant at $p<$0.1. }
\label{feti_teff}
\end{figure}

\begin{table*}
{\tiny
\renewcommand\arraystretch{1.0}
\caption{Input values of the atmospheric parameters and results of the spectroscopic analysis derived for the stars in our sample from the optical analysis. The derived abundances of Fe and Ti are also reported. The uncertainties on the abundances are $\sigma_1$ and $\sigma_2$, which are due to the EW measurements and related to the atmospheric parameters, respectively.  }
\setlength\tabcolsep{2.2pt}
\label{param}
\centering
\begin{threeparttable}
\begin{tabular}{lccccccccccc}
\toprule
ID &  $\rm{T_{eff,phot}}$ & \logg$_{\rm{trig}}$ & $\xi$ & $\rm{T_{eff,spec}}$ & \logg$_{\rm{spec}}$ & $\xi_{\rm{spec}}$ & [Fe/H]$\rm{_I}$ & [Fe/H]$\rm{_{II}}$ & [Ti/H]$\rm{_I}$ & [Ti/H]$\rm{_{II}}$\\
&\footnotesize{(K)} & \footnotesize{(dex)} & \footnotesize{(km\,s$^{-1}$)}&\footnotesize{(K)}&\footnotesize{(dex)}& \footnotesize{(km\,s$^{-1}$)}& \footnotesize{(dex)} & \footnotesize{(dex)}& \footnotesize{(dex)}& \footnotesize{(dex)} \\
\midrule
& & & & & &\bf{Standard stars} \\
HD3765&  5111$\pm$92 & 4.56$\pm$0.02 & 0.79$\pm$0.04 & 5001$\pm$75 & 4.56$\pm$0.10 & 0.59$\pm$0.20 & 0.02$\pm$0.01$\pm$0.08 & 0.13$\pm$0.03$\pm$0.09 & 0.14$\pm$0.01$\pm$0.11 & 0.13$\pm$0.02$\pm$0.06 \\
HD159222& 5863$\pm$111 & 4.41$\pm$0.05 & 1.03$\pm$0.04 & 5900$\pm$75 & 4.43$\pm$0.05 & 1.03$\pm$0.10 & 0.14$\pm$0.01$\pm$0.06 & 0.12$\pm$0.02$\pm$0.04 & 0.09$\pm$0.01$\pm$0.07 & 0.08$\pm$0.01$\pm$0.03\\
\hline
\\
& & & & & &\bf{Coma Berenices} \\
TYC\,1991-1235-1 &  5040$\pm$108 & 4.62$\pm$0.04 & 0.77$\pm$0.04 & 5070$\pm$70 & 4.53$\pm$0.05 & 0.87$\pm$0.12 & $-$0.05$\pm$0.01$\pm$0.04 & $-$0.03$\pm$0.03$\pm$0.07 & $-$0.03$\pm$0.01$\pm$0.09 & $-$0.02$\pm$0.02$\pm$0.04 \\
HIP\,61205 & 5972$\pm$114 & 4.58$\pm$0.02 & 1.05$\pm$0.04 & 5825$\pm$50 & 4.52$\pm$0.05 & 1.05$\pm$0.10 & $-$0.04$\pm$0.01$\pm$0.04 & $-$0.03$\pm$0.02$\pm$0.03 & $-$0.06$\pm$0.01$\pm$0.05 & $-$0.05$\pm$0.01$\pm$0.03  \\
TYC\,1989-0049-1 &  4718$\pm$76 & 4.73$\pm$0.02 & 0.67$\pm$0.04 & 4718$\pm$50 & 4.61$\pm$0.05 & 0.63$\pm$0.07 & $-$0.06$\pm$0.02$\pm$0.03 & 0.01$\pm$0.04$\pm$0.07 & 0.00$\pm$0.02$\pm$0.07 & 0.02$\pm$0.03$\pm$0.03  \\
TYC\,1989-147-1 &  5333$\pm$100 & 4.66$\pm$0.03 & 0.85$\pm$0.04 & 5313$\pm$50 & 4.51$\pm$0.05 & 0.80$\pm$0.10 & $-$0.05$\pm$0.01$\pm$0.04 & $-$0.01$\pm$0.03$\pm$0.05 & $-$0.04$\pm$0.01$\pm$0.06 & $-$0.02$\pm$0.02$\pm$0.04\\ 
\\
& & & & & &\bf{Ursa Major} \\
HD\,167389 &  6038$\pm$140 & 4.51$\pm$0.02 & 1.07$\pm$0.04 & 6000$\pm$75 & 4.55$\pm$0.05 & 1.07$\pm$0.15 & 0.01$\pm$0.01$\pm$0.06 & 0.00$\pm$0.02$\pm$0.04 & 0.03$\pm$0.01$\pm$0.07 & 0.02$\pm$0.02$\pm$0.04 \\
HD\,59747 &  5206$\pm$110 & 4.71$\pm$0.02 & 0.81$\pm$0.04 & 5106$\pm$50 & 4.58$\pm$0.05 & 0.73$\pm$0.10 & $-$0.02$\pm$0.01$\pm$0.03 & $-$0.01$\pm$0.03$\pm$0.05 & 0.01$\pm$0.01$\pm$0.06 & 0.02$\pm$0.02$\pm$0.03 \\
\\
& & & & & &\bf{Hercules Lyra} \\
HD\,70573 & 5755$\pm$129 & 4.43$\pm$0.03 & 1.00$\pm$0.04 & 5800$\pm$75 & 4.40$\pm$0.07 & 1.10$\pm$0.10 & $-$0.03$\pm$0.02$\pm$0.09 & 0.02$\pm$0.02$\pm$0.07 & $-$0.02$\pm$0.02$\pm$0.10 & 0.00$\pm$0.02$\pm$0.06 \\
\hline
\end{tabular}
\end{threeparttable}
}
\end{table*}

\begin{figure*}[!]
\centering
\includegraphics[width=1.\textwidth]{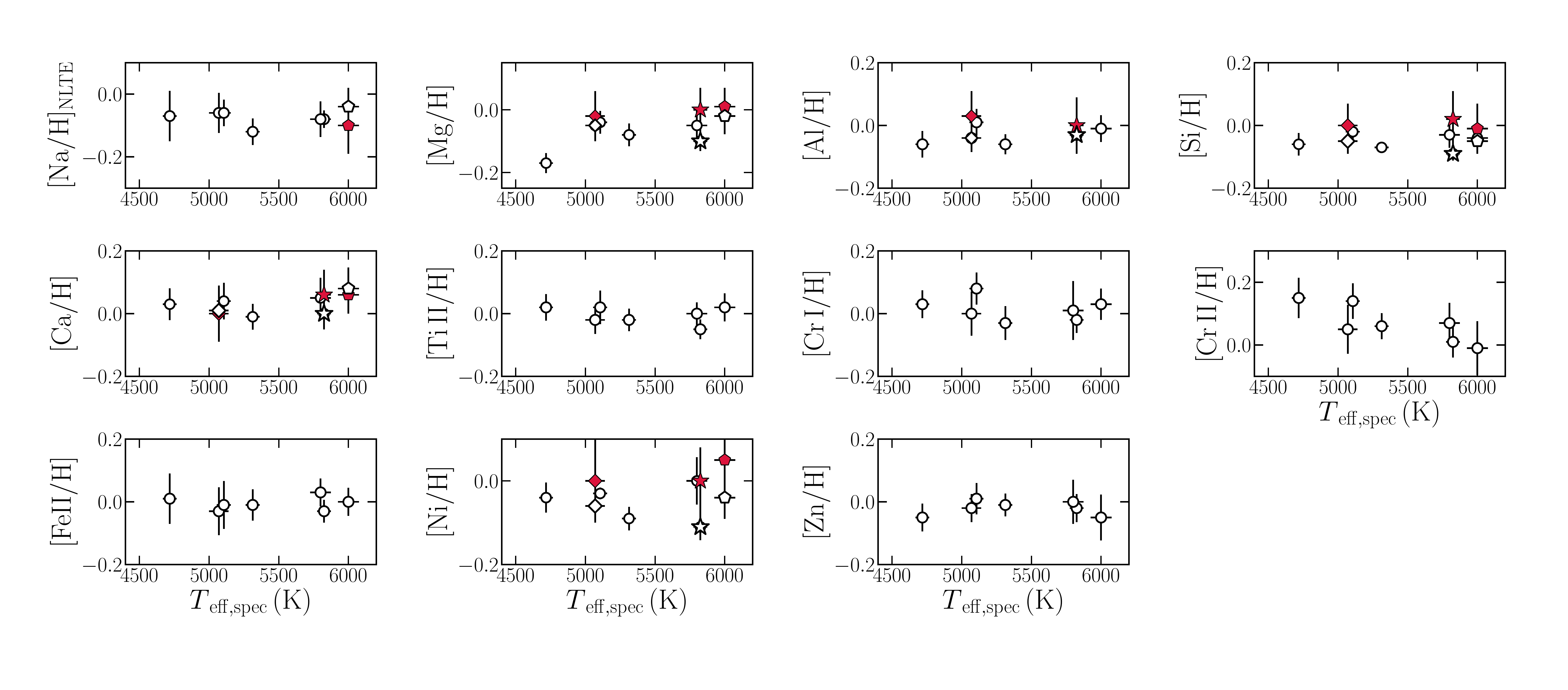}
\caption{Individual values of [X/H] as a function of spectroscopic estimates of \teff, derived from the analysis of the optical spectra (open symbols) and from the analysis of NIR spectra (red symbols). The symbols for the three stars for which we analysed GIARPS spectra are the same as in Fig. \ref{feti_teff}\,. All trends have Pearson correlation coefficients that are not significant at $p<$0.1, apart from Cr\,{\sc ii} (see the text for details). }
\label{other_teff}
\end{figure*}

\begin{sidewaystable}[!]
{\small
\caption{Abundance ratios obtained from the optical analysis. The two uncertainties in the abundances are $\sigma_1$ and $\sigma_2$, which are related to the EW measurements and atmospheric parameters, respectively.}
\centering
\setlength\tabcolsep{5pt}
\begin{tabular}{l|cc|cccc|cc|cc}
\toprule
$[\rm{X/Fe}]$ & HD3765$^{\ast}$ & HD159222 & TYC\,1991-1235-1 & HIP\,61205 & TYC\,1989-0049-1 & TYC\,1989-147-1 & HD\,167389 & HD\,59747 & HD\,70573\\
\midrule
$[\rm{C/Fe}]$ & $-$0.07$\pm$0.10$\pm$0.09 & $-$0.16$\pm$0.12$\pm$0.07 & $-$0.12$\pm$0.15$\pm$0.07 & $-$0.05$\pm$0.11$\pm$0.06 & $-$0.06$\pm$0.20$\pm$0.09 & $-$0.10$\pm$0.15$\pm$0.07 & 0.08$\pm$0.09$\pm$0.07 & $-$0.12$\pm$0.12$\pm$0.08 & -\\
$[\rm{Na/Fe}]_{\rm{NLTE}}$ & 0.11$\pm$0.01$\pm$0.07 & 0.00$\pm$0.02$\pm$0.03 & $-$0.05$\pm$0.04$\pm$0.04 & $-$0.04$\pm$0.02$\pm$0.06 & $-$0.15$\pm$0.07$\pm$0.08 & $-$0.12$\pm$0.03$\pm$0.14 & $-$0.05$\pm$0.02$\pm$0.14 & $-$0.08$\pm$0.03$\pm$0.06 & $-$0.02$\pm$0.04$\pm$0.03\\
$[\rm{Mg/Fe}]$ & 0.12$\pm$0.01$\pm$0.05 & $-$0.01$\pm$0.05$\pm$0.02 & 0.01$\pm$0.02$\pm$0.01 & $-$0.06$\pm$0.01$\pm$0.10 & $-$0.11$\pm$0.04$\pm$0.19 & $-$0.02$\pm$0.02$\pm$0.04 & $-$0.03$\pm$0.03$\pm$0.06 & $-$0.02$\pm$0.02$\pm$0.03 & $-$0.02$\pm$0.03$\pm$0.05\\
$[\rm{Al/Fe}]$ & 0.24$\pm$0.01$\pm$0.05 & 0.06$\pm$0.01$\pm$0.11 & 0.02$\pm$0.03$\pm$0.04 & 0.01$\pm$0.03$\pm$0.14 & 0.00$\pm$0.04$\pm$.03 & 0.00$\pm$0.01$\pm$0.01 & $-$0.03$\pm$0.03$\pm$0.06 & 0.03$\pm$0.03$\pm$0.06 & -\\
$[\rm{Si/Fe}]$ & 0.08$\pm$0.03$\pm$0.03 & $-$0.02$\pm$0.01$\pm$0.07 & 0.00$\pm$0.02$\pm$0.05 & $-$0.05$\pm$0.02$\pm$0.09 & 0.00$\pm$0.03$\pm$0.04 & $-$0.01$\pm$0.01$\pm$0.05 & $-$0.05$\pm$0.02$\pm$0.10 & 0.00$\pm$0.01$\pm$0.04 & 0.00$\pm$0.04$\pm$0.04\\
$[\rm{Ca/Fe}]$ & 0.14$\pm$0.02$\pm$0.07 & $-$0.01$\pm$0.03$\pm$0.02 & 0.06$\pm$0.01$\pm$0.12 & 0.04$\pm$0.03$\pm$0.07 & 0.09$\pm$0.02$\pm$0.17 & 0.04$\pm$0.01$\pm$0.08 & 0.06$\pm$0.03$\pm$0.10 & 0.07$\pm$0.03$\pm$0.13 & 0.08$\pm$0.04$\pm$0.13\\
$[\rm{Ti/Fe}]_I$ & - & $-$0.04$\pm$0.01$\pm$0.16 & 0.03$\pm$0.01$\pm$0.10 & $-$0.02$\pm$0.01$\pm$0.10 & 0.06$\pm$0.03$\pm$0.12 & 0.01$\pm$0.01$\pm$0.10 & 0.01$\pm$0.01$\pm$0.13 & 0.03$\pm$0.01$\pm$0.09 & 0.00$\pm$0.02$\pm$0.13\\
$[\rm{Ti/Fe}]_{II}$ & - & $-$0.04$\pm$0.02$\pm$0.07 & 0.01$\pm$0.06$\pm$0.09 & $-$0.03$\pm$0.02$\pm$0.08 & 0.01$\pm$0.04$\pm$0.09 & $-$0.01$\pm$0.04$\pm$0.07 & 0.02$\pm$0.03$\pm$0.06 & 0.03$\pm$0.04$\pm$0.15 & $-$0.03$\pm$0.03$\pm$0.08\\
$[\rm{Cr/Fe}]_I$ & 0.16$\pm$0.02$\pm$0.08 & $-$0.01$\pm$0.01$\pm$0.01 & 0.05$\pm$0.02$\pm$0.11 & 0.02$\pm$0.01$\pm$0.04 & 0.09$\pm$0.03$\pm$0.17 & 0.02$\pm$0.02$\pm$0.06 & 0.02$\pm$0.01$\pm$0.05 & 0.10$\pm$0.01$\pm$0.19 & 0.04$\pm$0.05$\pm$0.09\\
$[\rm{Cr/Fe}]_{II}$ & 0.34$\pm$0.02$\pm$0.08 & 0.01$\pm$0.04$\pm$0.02 & 0.07$\pm$0.06$\pm$0.14 & 0.04$\pm$0.04$\pm$0.06 & 0.14$\pm$0.06$\pm$0.25 & 0.07$\pm$0.03$\pm$0.12 & $-$0.01$\pm$0.07$\pm$0.04 & 0.14$\pm$0.04$\pm$0.27 & 0.04$\pm$0.05$\pm$0.06\\
$[\rm{Ni/Fe}]$ & 0.13$\pm$0.02$\pm$0.03 & $-$0.01$\pm$0.01$\pm$0.02 & 0.00$\pm$0.02$\pm$0.03 & $-$0.07$\pm$0.01$\pm$0.12 & 0.02$\pm$0.04$\pm$0.05 & $-$0.04$\pm$0.02$\pm$0.07 & $-$0.06$\pm$0.01$\pm$0.10 & $-$0.01$\pm$0.01$\pm$0.03 & 0.03$\pm$0.04$\pm$0.04\\
$[\rm{Zn/Fe}]$ & 0.04$\pm$0.03$\pm$0.05 & 0.00$\pm$0.01$\pm$0.05 & 0.03$\pm$0.02$\pm$0.06 & 0.03$\pm$0.02$\pm$0.05 & 0.01$\pm$0.05$\pm$0.03 & 0.04$\pm$0.02$\pm$0.06 & $-$0.06$\pm$0.02$\pm$0.15 & 0.04$\pm$0.04$\pm$0.05 & 0.03$\pm$0.05$\pm$0.03\\
\hline
\end{tabular}
\begin{tablenotes}
\small
\item $^{\ast}$ For HD\,3765, the [X/H] values are reported, since Fe does not satisfy the ionization equilibrium. The [Ti/H] values are reported in Table \ref{param}.
\end{tablenotes}
\label{ratios}
}
\end{sidewaystable}

\section{Discussion}

\subsection{Stellar parameters and elemental abundances}

The final values of the atmospheric parameters for all the stars analysed in our sample, as well as abundances of Fe\,{\sc i}, Fe\,{\sc ii}, Ti\,{\sc i,} and Ti\,{\sc ii}, are reported in Table \ref{param}.  All the abundance ratios obtained from the optical analysis are reported in Table \ref{ratios}, where the errors were derived as described in Sect. 3. Given the relatively wide range in \teff of our targets, we applied NLTE corrections to Na and C abundances derived from the optical range, following the prescriptions given by \cite{2011lind} and \cite{2019amarsi}, respectively. The final NIR abundances and uncertainties are reported in Table \ref{nir}.

The atmospheric parameters we derived with the new approach agree well with the input estimates used in the analysis, as shown in Fig. \ref{conf_param}. We calculated the mean difference between the initial guesses and the final spectroscopic values for each parameter. The temperatures are in excellent agreement; $\Delta T_{\rm{eff}}$=33$\pm$64\,K. For the comparison between the spectroscopic and trigonometric gravities, we find a mean difference of $-$0.08$\pm$0.06\,dex. As already noted by several authors \citep{2007sozzetti,2013tsantaki,2015maldonado}, the spectroscopic gravities tend to be under-estimated with respect to the trigonometric values, especially for \logg$>4.50$\,dex, where \teff$<5000$\,K. This is again a manifestation of the ionisation balance problem affecting cool dwarf stars, that is enhanced as the stellar age decreases. As an additional check, for the stars with significant difference between spectroscopic and trigonometric gravities the Gaia DR2 astrometric solutions\footnote{This work has made use of data from the European Space Agency (ESA) mission {\it Gaia} (\url{https://www.cosmos.esa.int/gaia}), processed by the {\it Gaia} Data Processing and Analysis Consortium (DPAC, \url{https://www.cosmos.esa.int/web/gaia/dpac/consortium}). Funding for the DPAC has been provided by national institutions, in particular the institutions participating in the {\it Gaia} Multilateral Agreement.} (including the parallax) are all well behaved, based on the reduced unit weight error (RUWE) metric (see e.g. \citealt{2018lindegren}). This further argues for the discrepancy arising because of above-mentioned limits in the spectroscopic measurements. Finally, regarding the $\xi$ values, we find a mean difference of $-0.004\pm$0.066\,\kms. 

In Figs.\ref{feti_teff} and \ref{other_teff}, we report our abundances as a function of \teff: the open symbols refer to the values we obtained from the optical analysis, while the red symbols represent the results from the NIR analysis. In particular, the diamond symbol is TYC\,1991-1235-1, the star symbol stands for HIP\,61205, and  the pentagon symbol represents HD\,167389. The lack of systematic trends between the derived optical abundances and \teff estimates validates our derivation of the atmospheric parameters. We calculated the Pearson correlation coefficient for all the trends in the two figures: none of these is statistically significant at $p$-value < 0.1 with the exception of Cr\,{\sc ii}. 
Despite the low [Mg/H] value obtained for TYC\,1989-0049-1, equal to $-$0.17$\pm$0.03$\pm$0.04, the trend [Mg/H] versus \teff has a Pearson correlation coefficient $r$=0.548, and $p$=0.2 . For this star only optical spectra are available, so we could not compare this low value with the NIR estimate. We only measured the abundance for the line 4730\AA\, for which NLTE corrections for the Sun are of the order of 0.01\,dex, as calculated by \cite{2016zhao}. The 5711\AA\, line is strong in the spectrum of this star, has an EW of 143\,m\AA,\, and according to our selection criteria, this line was excluded from the line list for the derivation of the abundance. According to \cite{2015osorio}, the NLTE corrections for a star such as TYC\,1989-0049-1 are very small, of the order of $-$0.006\,dex and we expect the NLTE corrections of the line 4730\AA\, are of the same order.  

The ionization equilibrium is satisfied for Ti and also for Fe for the stars in our sample, as shown in Table \ref{param}. Interestingly, this is not true for Cr, for which we find an anti-correlation with \teff. As shown in Fig. \ref{other_teff}, the Cr\,{\sc ii} abundances increase at decreasing \teff, especially for stars with \teff$\lesssim$5400\,K.  As already mentioned in Sec.\ref{introduction}, this can be explained by the over-ionisation effect. Differences between the neutral and ionised species for some atomic species, such as Fe, Ti, and Cr, have been observed in cool dwarfs with \teff$\lesssim 5400$\,K \citep{2000king,2007ramirez,2009dorazi,2010schuler}. These differences can reach values up to 0.6-0.8\,dex in stars younger than 100\,Myr and consequently the value of \logg should be decreased. The over-ionization effect is seen in cool dwarf stars, both in OC stars and in field stars \citep{2014bensby,2019tsantaki}. We can see this effect in the cool (\teff=5001\,K) standard star HD\,3765 (age $ \sim 5$\,Gyr). While the ionisation equilibrium is satisfied for Ti, suggesting a good estimate of \logg for this star, we find a large discrepancy between Fe\,{\sc i} and Fe\,{\sc ii} of about +0.11 dex, as already noted in \cite{2007ramirez}. A similar discrepancy is also seen for the Cr abundances, for which we obtained a difference of +0.18\,dex between Cr\,{\sc i} and Cr\,{\sc ii}, as shown in Table \ref{ratios}. In the analysis, we used two Cr\,{\sc ii} lines that are not blended with known contaminants, according to \cite{2017lawler}. We note that the NIR Fe abundance is in agreement with the Fe\,{\sc ii} estimate in the optical range. The reason for such observed discrepancies is still unknown: this may be due to the limitations of 1D-LTE model atmospheres, 3D effects, stellar activity, or a combination of these.

The agreement between the optical and NIR abundances for the stars is overall good,  within the uncertainties. Our results corroborate the previous findings of \cite{2019caffau}, who derived chemical abundances for different species of 40 stars by analysing  spectra from GIANO in its previous configuration (fibre-fed).  However, we noted that for the star HIP\,61205 we obtained larger discrepancies for Mg, Si and Ni between the optical and NIR abundance; the latter values are nearly solar. Such discrepancies could be related to the different number of lines used to derive the abundances: more in the optical range (16 lines) than in the NIR (1 line).

\begin{figure*}[!] 
\centering
\includegraphics[width=0.9\textwidth]{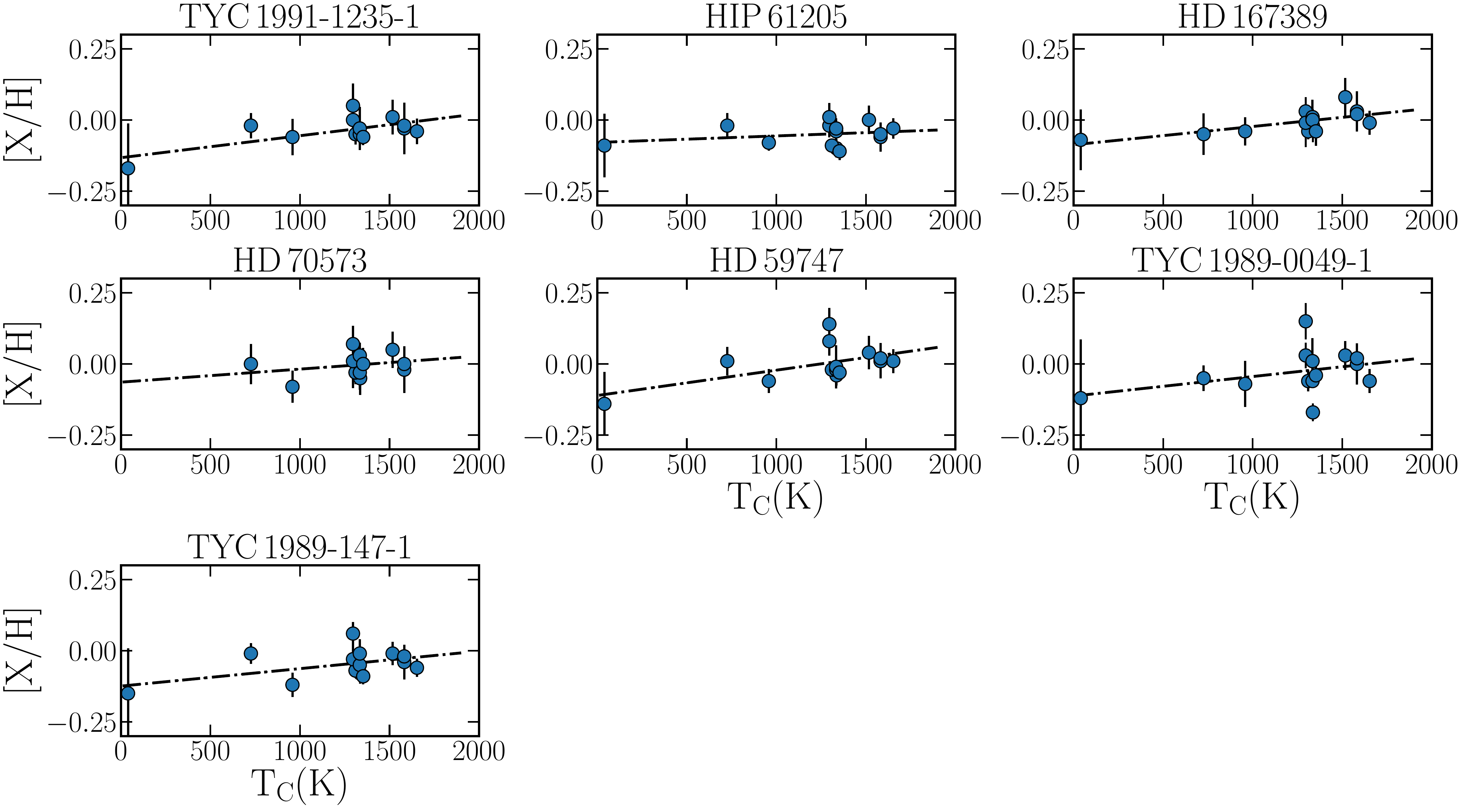}
\caption{Values of [X/H] as a function of the condensation temperature T$_{\rm{C}}$, taken from \cite{2003lodders}.  }
\label{xh_tcond}
\end{figure*}

In Fig. \ref{xh_tcond}, the abundance ratios [X/H] are plotted as a function of the condensation temperature T$_{\rm{C}}$, taken from \cite{2003lodders}. Given the wide range in \teff covered by the stars in our sample ($\sim$1500\,K), we could not perform a strictly differential analysis with respect to stars of the same association and with similar \teff (see e.g. \citealt{2009melendez,2014melendez}). For each trend we calculated the Pearson and Spearman correlation coefficients and we find that the trend is significant at $p < $0.05 in both cases  for HD\,167389 alone. It has been suggested that the positive slopes observed in [X/H]-T$_{\rm{C}}$ plots might be the result of accretion onto the star of refractory material present in the circumstellar disc or a signature of planet engulfment episodes. Instead, for the other stars, the correlation coefficients are not in agreement; thus no exhaustive conclusions can be drawn in those cases.

\subsection{Comparison with previous studies}

\begin{table*}
{\small
\centering  
\caption{Mean values of the [X/H] ratios derived from the NIR analysis for four stars in our sample. }
\setlength\tabcolsep{4.5pt}
\centering
\begin{threeparttable}
\begin{tabular}{lccccccc}
\toprule
$\rm{[X/H]}$ & HD3765 & HD159222 & TYC\,1991-1235-1 & HIP\,61205 & HD\,167389\\
\midrule
C & 0.11$\pm$0.10$\pm$0.08 & 0.18$\pm$0.09$\pm$0.06 & $-$0.03$\pm$0.15$\pm$0.09 & 0.02$\pm$0.12$\pm$0.06 & -\\
Na & - &        0.20$\pm$0.08$\pm$0.06 & - & - & $-$0.10$\pm$0.10$\pm$0.07\\
Mg & 0.15$\pm$0.02$\pm$0.07 &   0.13$\pm$0.03$\pm$0.06 & $-$0.02$\pm$0.01$\pm$0.08 & $-$0.03$\pm$0.03$\pm$0.06 & 0.01$\pm$0.03$\pm$0.07\\
Al & 0.25$\pm$0.08$\pm$0.09 &   0.22$\pm$0.10$\pm$0.08 & 0.03$\pm$0.12$\pm$0.08 & 0.00$\pm$0.12$\pm$0.08 &      -\\
Si & 0.12$\pm$0.03$\pm$0.06  &  0.04$\pm$0.04$\pm$0.05 & 0.00$\pm$0.03$\pm$0.06 & 0.02$\pm$0.02$\pm$0.06 & $-$0.01$\pm$0.01$\pm$0.05\\
Ca & 0.16$\pm$0.07$\pm$0.07 &   0.10$\pm$0.03$\pm$0.04 & 0.02$\pm$0.02$\pm$0.09 & 0.06$\pm$0.09$\pm$0.05 & 0.08$\pm$0.03$\pm$0.06\\
Ti & 0.15$\pm$0.10$\pm$0.06 &   0.09$\pm$0.09$\pm$0.05 & 0.00$\pm$0.10$\pm$0.07 & 0.05$\pm$0.08$\pm$0.06 & $-$0.05$\pm$0.07$\pm$0.06\\
Fe & 0.11$\pm$0.05$\pm$0.08 &   0.17$\pm$0.09$\pm$0.07 & $-$0.04$\pm$0.06$\pm$0.07 & $-$0.02$\pm$0.04$\pm$0.06 & $-$0.01$\pm$0.07$\pm$0.06\\
Ni & 0.15$\pm$0.06$\pm$0.08 &   0.20$\pm$0.08$\pm$0.06 & 0.00$\pm$0.12$\pm$0.06& 0.00$\pm$0.09$\pm$0.07 & 0.05$\pm$0.11$\pm$0.05\\
\hline
\end{tabular}
\end{threeparttable}
\label{nir}
}
\end{table*}

Our abundance measurements are in overall fair agreement with other studies found in the literature, as shown in Table \ref{comparison}, where we reported the mean values for each cluster and results from different studies. Regarding the Coma Berenices cluster, we find that our measurements are in good agreement with \cite{2015blanco-cuaresma} for all the atomic species. \cite{2016netopil} report a mean [Fe/H] equal to 0.00$\pm$0.08, averaging over different estimates in the literature. Other studies on the chemical composition of this cluster are mainly focussed on the analysis of A-F type stars that have temperatures higher than 6000\,K. \cite{2000burkhart} reported a mean [Fe/H]=$-0.07\pm0.05$\,dex for 1 F-type star; on the contrary, \cite{2008gebran} found $<\rm{[Fe/H]}>$=0.07$\pm$0.09\,dex for 11 F-type stars analysed through the spectral synthesis technique. The large difference between these two studies could be related to the different line lists and techniques employed and also to the different number of stars analysed. \cite{1992friel} analysed high-resolution, high S/N spectra of 14 F-G type stars, with \teff$>5950$\,K, through the EW method. In particular, they analysed the spectral window 6078-7755\,\AA\, , where they measured 8 Fe\,{\sc i} lines. These authors found a mean metallicity of $-$0.05$\pm$0.05\,dex, again in good agreement with our measurements, despite the different type of stars analysed. Regarding the individual stars, we find that our measurement for HIP\,61205 confirmed the results of \cite{2016brewer}. The authors find \teff=5796\,K, \logg=4.51\,dex and [Fe/H]=$-$0.02, which are in excellent agreement with our estimates. \cite{2016brewer} also derived abundances for different atomic species, in particular they find that [C/Fe]=$-$0.04, [Na/Fe]=$-$0.12, [Mg/Fe]=$-$0.06, [Al/Fe]=$-$0.12, [Si/Fe]=$-$0.04, [Ca/Fe]=0.02, [Ti/Fe]=$-$0.02, [Cr/Fe]=0.01, and [Ni/Fe]=$-$0.09. All abundance ratios agree very well with our estimates, as shown in Table \ref{ratios}. To our knowledge, for TYC\,1991-1235-1, TYC\,1989-0049-1, and TYC\,1989-147-1, there are no previous studies on abundances in the literature.

For the UMa moving group, our estimate of mean [Fe/H]=$-$0.01$\pm$0.01 is in fair agreement with the results from \cite{1993soderblom}, \cite{2005kingschuler}, and \cite{2005monier}, which reported mean values equal to $-$0.08$\pm$0.09, $-$0.06$\pm$0.05, and $-$0.05$\pm$0.02, respectively. Our results also confirm what \cite{2012biazzo} and \cite{2017tabernero} find; these authors analysed stars similar to our sample employing the EW method. As shown in Table \ref{info_let}, our results agree well with the two studies. HD\,167389 has been analysed by \cite{2009ammler} and \cite{2017tabernero}. In particular, the former derived the stellar parameters and abundances of Fe and Mg through spectral synthesis: they find \teff=5895$\pm$80\,K, \logg=4.37$\pm$0.15\,dex, $\xi$=0.99$\pm$0.20\,\kms, [Fe/H]=$-$0.02$\pm$0.07, [Mg/Fe]=$-$0.03$\pm$0.05, in excellent agreement with our estimates. \cite{2017tabernero} analysed candidate members of the UMa group to confirm their membership through chemical tagging by employing the EW analysis method. The authors find for HD\,167389 \teff=5978\,K, \logg=4.56\,dex and [Fe/H]=+0.01,  confirming our results. Moreover, they derived abundances for various atomic species, finding [Na/Fe]=$-$0.06$\pm$0.01, [Mg/Fe]=$-$0.07$\pm$0.03, [Al/Fe]=$-$0.05$\pm$0.01, [Si/Fe]=$-$0.02$\pm$0.01, [Ca/Fe]=0.03$\pm$0.01, [Ti/Fe]=0.01$\pm$0.01, [Cr/Fe]=$-$0.01$\pm$0.01, and [Ni/Fe]=$-$0.04$\pm$0.01. HD\,59747 was analysed by \cite{2009ammler}, who derived the stellar parameters, [Fe/H], and [Mg/H] through spectral synthesis fitting. The authors found \teff=5094\,K, \logg=4.55\,dex, [Fe/H]=$-$0.03, and [Mg/Fe]=$-$0.01, in excellent agreement with our results. 

In the Her-Lyr association, we analysed only the star HD\,70573, which is also the star with the highest \vsini\,\,in our sample.  This star was also analysed by \cite{2010gonzalez} in the standard way, that is using Fe (neutral and ionised) lines to derive the atmospheric parameters. These authors found \teff=5807$\pm$85\,K, \logg=4.35$\pm$0.08\,dex, $\xi$=1.80$\pm$0.16\,\kms and [Fe/H]=$-$0.05$\pm$0.06.
These results are also confirmed by \cite{2010ghezzi}, who analysed the star in the same way, finding \teff=5884$\pm$26\,K, \logg=4.57$\pm$0.08\,dex, $\xi$=1.69$\pm$0.06\,\kms and [Fe/H]=$-$0.04$\pm$0.03. Our results confirm the values reported in the two different studies, with the exception of $\xi$, for which we find a lower value equal to 1.10$\pm$0.10 \kms; there is a difference of the order of 0.6-0.7\,\kms between the two studies. Such a discrepancy could be due to the different line lists used, in particular to differences in the atomic data, but they also seem to confirm the results of \cite{2020baratella}, regarding the possible overestimation of the $\xi$ parameter when using iron lines.
As shown in Table \ref{comparison}, our results confirm those found by \cite{2016brewer}. However we note large differences for some elements, such as Na, Mg, and Ni, which could be due to differences in the line list used, in particular to differences in the atomic line parameters, and to the different spectroscopic analysis technique employed. 
\\

\subsection{Carbon abundances}

We measured C abundance using four different indicators: two high-excitation potential lines at 5380.337 and 6587.61 \AA\, (atomic data and NLTE corrections from \citealt{2019amarsi}); the NIR line at 16021.7 \AA\,(atomic data from \citealt{2015shetrone}); and the CH molecular band at 4300\AA\,, for which molecular line data come from Plez (priv. communication). We note that the line 16021\AA\, suffers from a blend with Fe and Si lines in the blue wing. This blend is not significant for the Sun, TYC\,1991-1235-1, and HIP\,61205, but it becomes more important in HD\,167389 (\teff=6000\,K), because the Si feature is a high-excitation line ($\chi=7.035$\,eV), which strengthens at these temperatures. For this reason, we could not derive the abundance in the NIR range for this star. The abundances of C\,{\sc i} lines in the optical part were calculated with the EW method, while the abundances for the NIR lines and for the CH feature were calculated through spectral synthesis. We applied NLTE corrections to C\,{\sc i} optical abundances following \cite{2019amarsi}. Despite the wide range in \teff covered by the stars we analysed, NLTE corrections of optical lines are small, typically of the order of $-$0.01\,dex.

Carbon is one of the most important elements for life on Earth and also for planetary formation models. In particular, deriving reliable estimates of the C/O ratio is crucial, since it provides clues to where the planets formed in the protoplanetary disc and possible subsequent radial migration \citep{2017brewer}.
However, we could not derive abundances of oxygen (O) from OH molecular features in the NIR, because in F-G stars molecular absorption is less important, weakening the lines \citep{2018souto}. For the K-type star TYC\,1989-0049-1 GIANO-B spectrum is not available, hampering the determination of abundances through the NIR molecular lines. Also, extremely high-resolution, high S/N spectra are required to be able to measure OH lines \citep{2004melendez}. Additionally, the HARPS-N spectra does not allow us to cover the O I triplet at 7773\,\angstrom, which is ideal for solar-type stars. Despite the forest of CN lines in the solar spectrum covering the blue, red, and NIR part of it \citep{1982sneden}, we did not derive N abundance from those molecular lines. First of all, for solar-type stars the best tools to derive reliable N abundance are high-excitation N\,{\sc i} atomic lines at 7400-8720\,\angstrom\  \citep{asplund,2014sneden}, which is not covered by HARPS-N (and GIANO-B) spectra. Moreover, we notice that N has little impact on the molecular equilibrium. Thus, we derive only C abundance, fixing O and N abundances to solar values, which is a reasonable assumption for our sample that is comprised of intermediate-age, thin disc, main-sequence stars (e.g. \citealt{2014bensby}).

\begin{figure}[!] 
\centering
\includegraphics[width=0.4\textwidth]{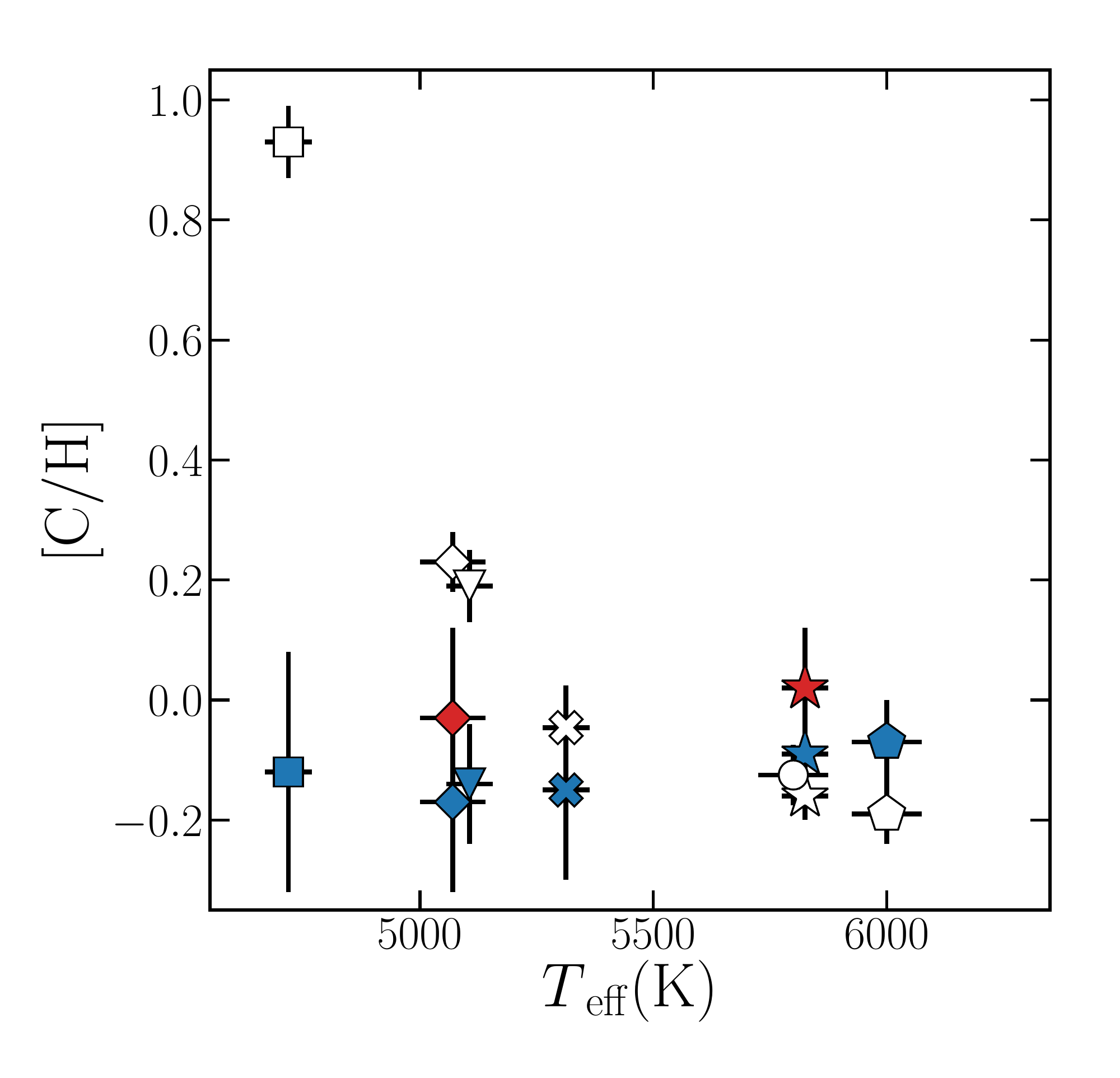}
\caption{Abundances of C\,{\sc i} as a function of \teff, derived from the optical analysis (empty symbols), from the NIR line (red symbols), and from CH band at 4300\AA\,. The different symbols are the different estimates for the same star: the diamond represents TYC\,1991-1235-1, the star HIP\,61205, the pentagon HD\,167389, the circle HD\,70573, the triangle HD\,59747, the square TYC\,1989-0049-1, and finally the x-shaped symbol indicates TYC\,1989-147-1.  }
\label{over_C}
\end{figure}

The values for the Sun are reported in Table \ref{solar}; for C abundance inferred from the CH features we obtained $\log_{\rm{n}}$(C)$_{\odot}$=8.35$\pm$0.08, that is marginally lower than that obtained from the atomic lines. Such difference can be becasue the atomic data of the CH feature are not so precise. 
In Fig.\ref{over_C} the different C abundance estimates are reported as a function of \teff. The empty symbols represent the values obtained in the optical range, the red symbols stand for the NIR measurements, and the blue symbols indicate the C values from CH molecular band. The different stars are represented by the different symbols, as described in the caption of the figure.
As shown, we obtained different trends from the different lines. The increasing C\,{\sc i} abundances at decreasing \teff for the optical measurements is noteworthy. However, this is not seen for the NIR abundances and the values derived from the CH. Since NLTE corrections of C\,{\sc i} optical abundances are negligible for the stars in our sample, we believe that its trend with \teff is due to over-excitation effects. \cite{2015schuler} find  similar behaviour when deriving C abundance from two high-excitation lines, which have $\chi$ similar to our adopted lines, for a star with \teff=5406\,K. The authors find that the C abundance inferred from atomic lines is +0.16\,dex higher than that derived from the C$_2$ feature. The interesting aspect is that even if the NIR C line has a high-excitation energy ($\chi$=9.631 eV), we do not see the same effect in the abundances. A possible explanation is that at 1.6\,$\mu$m there is the H$^-$ absorption minimum, so we see the deepest photospheric layers of the atmosphere where LTE is a good approximation to compute the populations of atomic levels.  \cite{2015schuler} argue that the over-excitation could be explained as a NLTE effect, specifically it could be the result of our incapacity to properly model the population of high-energy levels under LTE approximation.  Thus, the C abundances obtained from the NIR line, even if it has a high-excitation energy, could be the real C abundances of the stars. Moreover, while for star TYC\,1991-1235-1, \teff=5070\,K, we obtained a difference between optical and NIR abundances of +0.26\,dex, for HIP\,61205, \teff=5825\,K, the behaviour is reversed; in this star,\ the optical C abundanceis smaller than the NIR estimate by $-$0.22\,dex. Unfortunately, we could not confirm this trend in the whole temperature range, since GIANO-B spectra are not available for all the stars.  As a further test, we measured C abundance from the CH band at 4300\AA\,. We find that the trend with \teff is not statistically meaningful (with a $p$-value > 0.1) and, most importantly, the over-excitation effect observed for the C\,{\sc i} abundances is not present. Also, especially for TYC\,1991-1235-1, the C abundance from CH is in better agreement with the NIR estimate than the optical. We suggest that the values obtained from molecular features in the optical for very young stars are more reliable estimates of C abundances, as already suggested by \cite{2015schuler}. For this reason, the [C/Fe] values in Table \ref{ratios} are calculated with the C abundance derived from CH lines.

\begin{table*}
{\tiny
\renewcommand\arraystretch{1.0}
\caption{Mean values of the abundances ratios for each cluster and comparison with literature studies. For Hercules Lyra association only the star HD\,70573 was analysed, for which the errors are calculated as the quadratic sum of the $\sigma_1$ and $\sigma_2$ contributions.   }
\setlength\tabcolsep{2pt}\label{info_let}
\label{comparison}
\centering
\begin{threeparttable}
\begin{tabular}{lcccccccccccccc}
\toprule
Ref. &  $<[\rm{Fe/H}]>$ & $<[\rm{C/Fe}]>$ & $<[\rm{Na/Fe}]>$ & $<[\rm{Mg/Fe}]>$ & $<[\rm{Al/Fe}]>$ & $<[\rm{Si/Fe}]>$ & $<[\rm{Ca/Fe}]>$ & $<[\rm{Ti/Fe]}>$ &  $<[\rm{Cr/Fe]}>$ &  $<[\rm{Ni/Fe}]>$ &  $<[\rm{Zn/Fe}]>$\\
\midrule
&&&&&\textbf{Coma Berenices}\\
This work & $-$0.05$\pm$0.01 & $-$0.08$\pm$0.02 & $-$0.09$\pm$0.03 & $-$0.05$\pm$0.03 & 0.01$\pm$0.01 & $-$0.02$\pm$0.01 & 0.06$\pm$0.01 & 0.02$\pm$0.02 & 0.05$\pm$0.02 & $-$0.02$\pm$0.02 & 0.03$\pm$0.01\\
BC15$^a$ & $-$0.07$\pm$0.02 & - & $-$0.05$\pm$0.02 & 0.01$\pm$0.02 & - & 0.04$\pm$0.02 & 0.04$\pm$0.01 & 0.02$\pm$0.02 & 0.02$\pm$0.06 & $-$0.08$\pm$0.01 & -\\
\hline
&&&&&\textbf{Ursa Major}\\
This work & $-$0.01$\pm$0.02 & $-$0.02$\pm$0.07 & $-$0.07$\pm$0.02 & $-$0.03$\pm$0.01 & 0.00$\pm$0.03 & $-$0.03$\pm$0.03 & 0.07$\pm$0.01 & 0.02$\pm$0.01 & 0.06$\pm$0.04 & $-$0.03$\pm$0.02 & $-$0.01$\pm$0.05\\
T17$^b$ & 0.03$\pm$0.07 & - & $-$0.06$\pm$0.05 & $-$0.03$\pm$0.06 & $-$0.02$\pm$0.06 & 0.00$\pm$0.03 & 0.03$\pm$0.03 & 0.05$\pm$0.06 & 0.01$\pm$0.03 & $-$0.04$\pm$0.03 & -\\
B12$^c$ &  0.01$\pm$0.01 & - & $-$0.08$\pm$0.03 & 0.01$\pm$0.04 & 0.09$\pm$0.01 & $-$0.03$\pm$0.06 & 0.07$\pm$0.01 & 0.02$\pm$0.09 & 0.01$\pm$0.03 & $-$0.05$\pm$0.01 & $-$0.12$\pm$0.05 \\
\hline
&&&&&\textbf{Hercules Lyra}\\
This work & 0.00$\pm$0.01 & - & $-$0.02$\pm$0.05 & $-$0.02$\pm$0.06 & - & 0.00$\pm$0.06 & 0.08$\pm$0.14 & $-$0.01$\pm$0.01 & 0.04$\pm$0.01 & 0.03$\pm$0.05 & 0.03$\pm$0.06 \\
B16$^d$ & 0.08 & $-$0.05 & $-$0.14 & $-$0.14 & $-$0.23 & $-$0.08 & 0.05 & 0.00 & 0.03 & $-$0.14 & - \\
\hline
\end{tabular}
\begin{tablenotes}
\small{
\item $^a$ \cite{2015blanco-cuaresma}
\item $^b$ \cite{2017tabernero}
\item $^c$ \cite{2012biazzo}
\item $^d$ \cite{2016brewer}
}
\end{tablenotes}
\end{threeparttable}
}
\end{table*}

\subsection{Effects of stellar activity}

The over-excitation and over-ionisation effects that we observed for the C and Cr\,{\sc ii} abundances are among the principal problems affecting the analysis of young cool stars. Such effects are more evident in young, intermediate-age ($\tau \lesssim$ 800\,Myr) and cool dwarf stars (\teff$\lesssim$5400\,K), which are more intense at decreasing ages and temperatures. Young stars are more active and they have more intense chromospheric or photospheric magnetic fields than older stars. The main effect of local magnetic fields on spectral lines is the broadening of their profile through the Zeeman effect that causes a splitting of the spectral line into its multiplet components. This effect is directly proportional to the wavelength and to the value of the Landè $g_{\rm{L}}$ factor. The latter parameter measures the sensitivity of an atomic transition to the magnetic fields, meaning the higher the $g_{\rm{L}}$ factor, the more sensitive the line is to Zeeman splitting. The C\,{\sc i} NIR (16021\AA\,) and optical (5380\AA\, and 6588\AA\,) lines have a Landè factor equal to 1.15, 1.0, and 1.0, respectively. For the Cr\,{\sc ii} lines at 4848.23\,\AA\, and 5237.33\,\AA\, the $g_{\rm{L}}$ is equal to 1.25 and 1.30, respectively. According to \cite{2015shchukina}, these lines are insensitive to the presence of magnetic fields, which produce the stronger effects in lines with $g_{\rm{L}} \sim 2.0$ typically. Moreover, the Zeeman splitting has two main effects on the spectral line. On one hand, it produces a broadening of the profile and an increase of the EW. On the other hand, the line weakens, with a decrease of its depth; thus, the two effects compensate for each other \citep{2013reiners}. In this case, we can exclude the Zeeman effect as a possible explanation of the observed trends in Figs.\ref{other_teff} and \ref{over_C} . 

\begin{figure}[!] 
\centering
\includegraphics[width=0.4\textwidth]{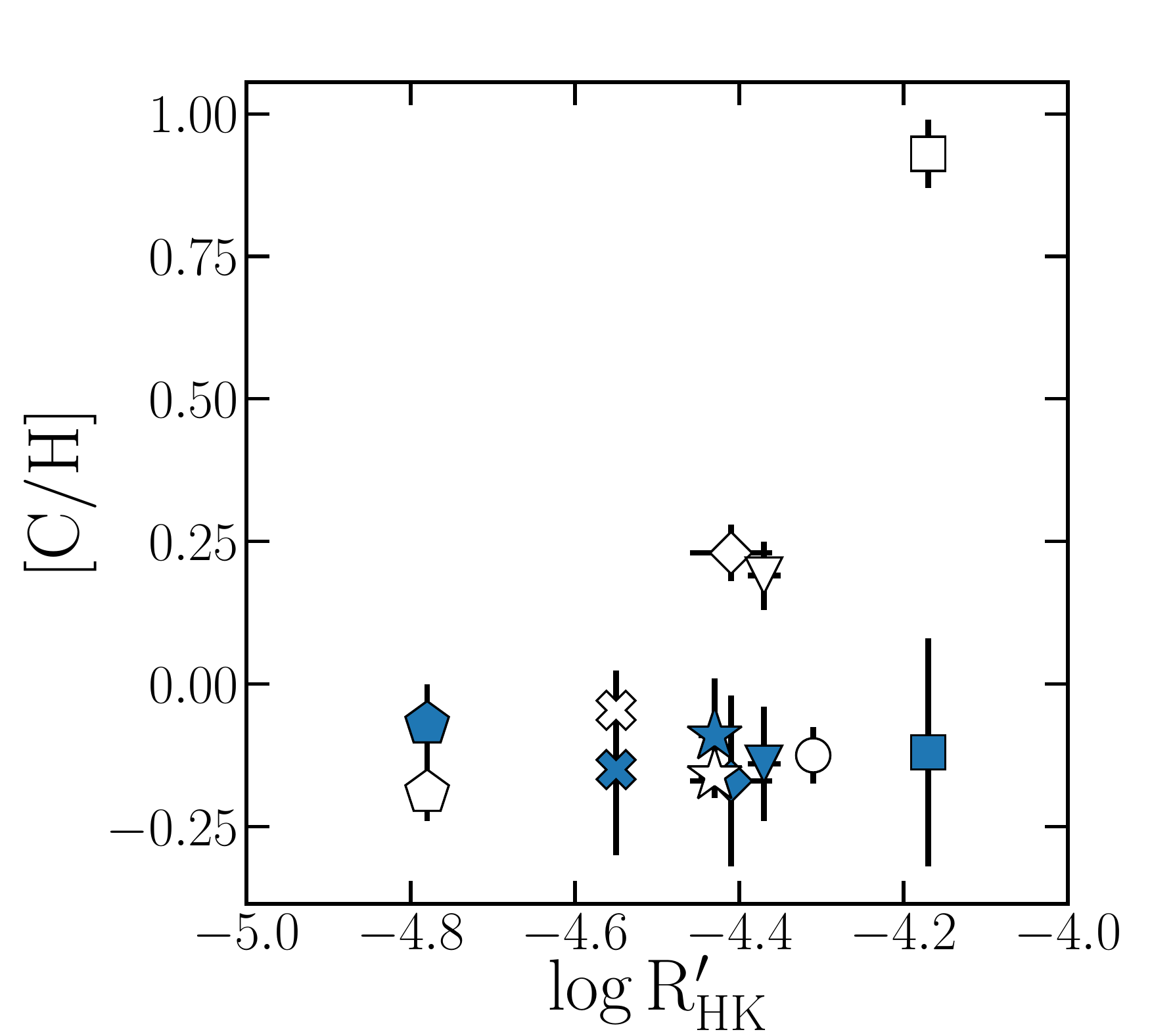}
\caption{Abundances of C\,{\sc i} derived from atomic lines in the optical range (empty symbols) and from the CH molecular features (blue symbols) as a function of the activity index $\log$R$^{\prime}_{\rm{HK}}$. The symbols are the same as in Fig.\ref{over_C}.  }
\label{C_activity}
\end{figure}

\begin{figure}[!] 
\centering
\includegraphics[width=0.4\textwidth]{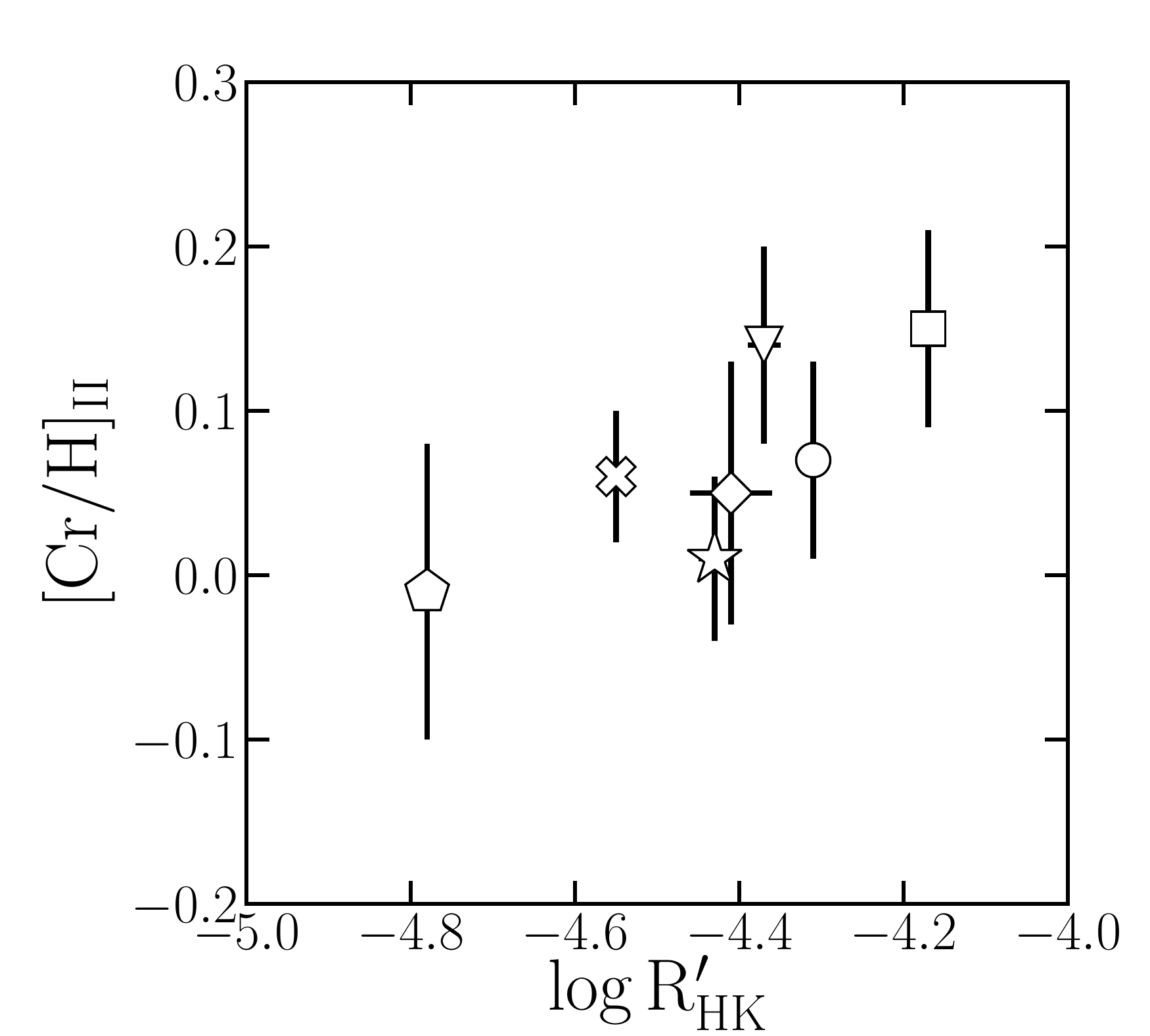}
\caption{Abundances of Cr\,{\sc ii} as a function of activity index  $\log$R$^{\prime}_{\rm{HK}}$. The symbols are the same as in Fig.\ref{over_C}.  }
\label{cr_activity}
\end{figure}

In Fig.\ref{C_activity}, we plot the estimates of the C abundances from the two atomic lines in the optical range and from the CH molecule as a function of the activity index $\log$R$^{\prime}_{\rm{HK}}$. The latter values were calculated with Yabi\footnote{\url{https://www.ia2.inaf.it}} interface \citep{PMID:22333270}, following the prescription of \cite{1984noyes} and through the procedure described by \cite{2011lovis}. Yabi is a Python web application installed at IA2 in Trieste that allows authorised users to run the HARPS-N DRS pipeline on proprietary data with custom input parameters. Since we analysed co-added spectra of the stars, we calculated the mean values of the activity indexes, averaging over the spectra we used in the co-adding procedure. The $\log$R$^{\prime}_{\rm{HK}}$ indexes are reported in Table \ref{tabinfo}. As shown, the C\,{\sc i} abundances from atomic lines in the optical range have a positive correlation with the $\log$R$^{\prime}_{\rm{HK}}$, with a Pearson correlation coefficient of $r$=0.69 and $p$-value=0.08. On the contrary, the abundance values derived from the CH features do not correlate with the activity indexes, again suggesting that these values are more reliable estimates of C abundances. We also find a similar behaviour for Cr\,{\sc ii} abundances, as shown in Fig.\ref{cr_activity}. We may envisage different, plausible scenarios to explain this peculiar trend. The effect of the chromospheric emission in active stars is observed mainly in the Lyman-$\alpha$ lines of H atom, the Ca\,{\sc ii} H and K lines, Mg\,{\sc ii} lines, and He lines. In particular, the photons from the Lyman-$\alpha$, with an energy of 10.2\,eV, can ionise Cr atoms that have a first ionisation potential of 6.77\,eV. So, the population of Cr\,{\sc ii} atoms is larger than Cr\,{\sc i} and this can qualitatively explain the increase of abundance at increasing levels of activity. In the case of C\,{\sc i} lines, a possible explanation of the overabundance is the presence of unknown blends in the optical lines that become stronger as the \teff decreases. Moreover, these blends could be more important in active stars than in quiet stars, if a significant part of the flux comes from cool regions, such as photospheric dark spots, where the lines are stronger and/or more sensible to the temperature. Another possible explanation is that the increase of the population of the atomic levels (from which the 5380 and 6587\,\AA\, lines form) is mainly due to UV continuum photons between 1450 \,\AA\, and 1650 \,\AA,\, which increase in intensity at increasing levels of activity \citep{2012linsky}. These photons might be responsible of the larger population of the levels from which the 5380\,\AA\, and 6587\,\AA\, lines are formed. Although we do not have a definitive explanation to the over-ionisation/excitation effects, the solutions proposed seem intriguing and worthy of a detailed investigation.  

\begin{figure}[!] 
\centering
\includegraphics[width=0.5\textwidth]{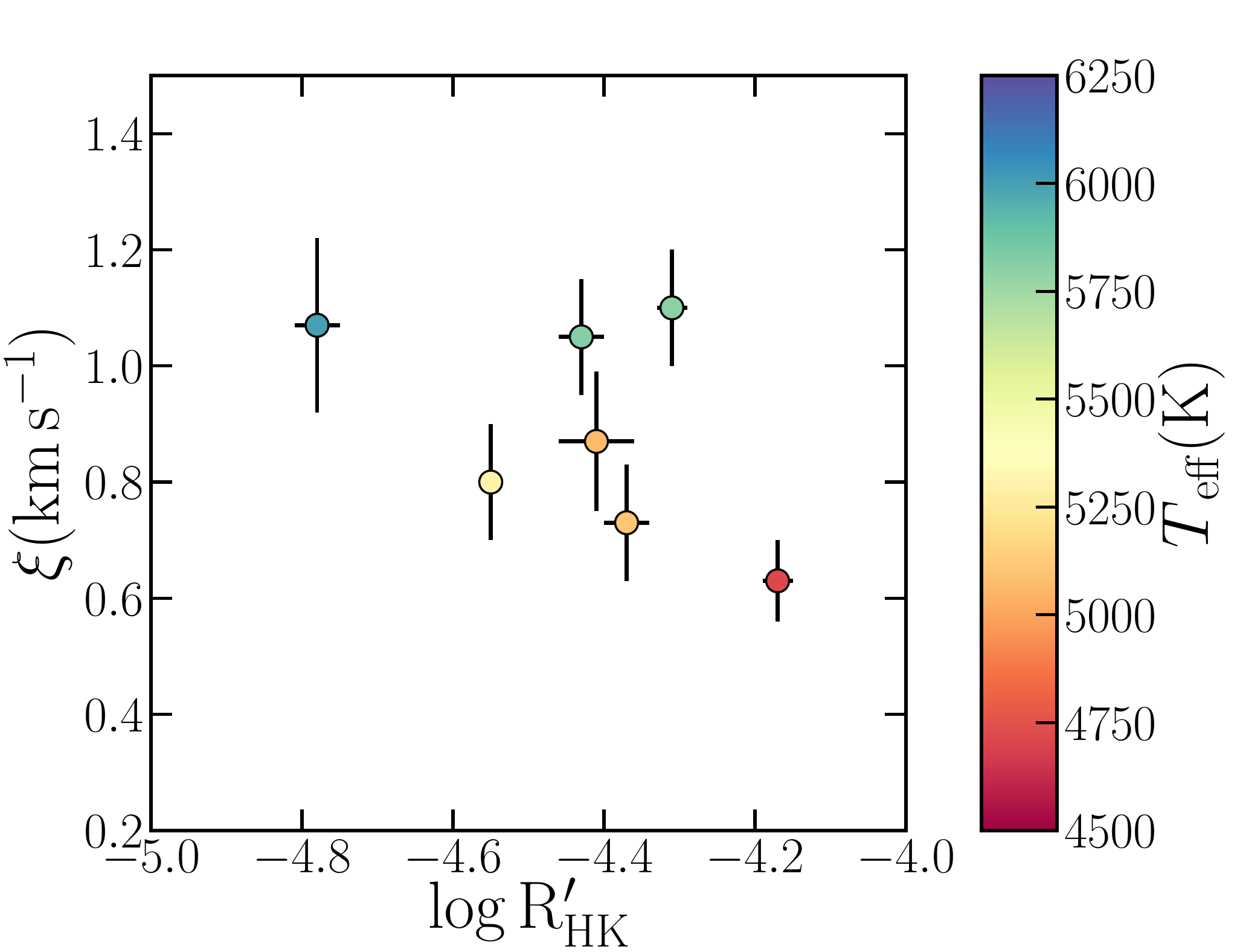}
\caption{Values of $\xi$ parameter derived with the new approach as a function of the chromospheric activity index $\log$R$^{\prime}_{\rm{HK}}$. The symbols are colour-coded according to the \teff. }
\label{xi_activity}
\end{figure}

We also find that the $\xi$ values we obtained using Ti lines do not seem to correlate with $\log $R$^{\prime}_{\rm{HK}}$, as shown in Fig.\ref{xi_activity}. The symbols in this figure are colour coded according to the \teff. As already known, the $\xi$ velocity increases systematically towards higher \teff and lower \logg. In particular, in dwarf stars (\logg$\sim 4.50$\,dex) the $\xi$ values are of the order of 0.70\,\kms at \teff$\sim$4500\,K \citep{2013steffen}. Thus, the trend of $\xi$ with \teff observed in Fig.\ref{xi_activity} is expected.   We calculated the Pearson correlation coefficient for the trend, that is equal to $r$=-0.49, with $p$=0.26; thus it is not significant at $p$<0.10. This result further validates our method and what was previously found by \cite{2020baratella}.

\section{Conclusions}

In this first paper of a series, we presented the preliminary results of an extensive analysis of optical and NIR spectra of stars observed by the GAPS-YO programme. In particular, we derived the atmospheric parameters and the chemical composition of seven target stars, the Sun, and two RV standard stars, HD159222 and HD3765, using a new spectroscopic approach to overcome analytical issues related to the relatively young ages of the stars.

The analysis of young and intermediate-age stars, in particular in the cool temperature regime (\teff $\lesssim$ 5400\,K), is not trivial, owing to a series of effects still unexplained from a theoretical point of view. For these reasons, we applied the same methodology as in \cite{2020baratella} for the analysis of the optical HARPS-N spectra. In general, our derived spectroscopic estimates of the atmospheric parameters are in excellent agreement with the initial guesses. The atmospheric parameters we obtained from the optical analysis were used to derive the abundances in the NIR part, through the spectral synthesis technique and using the same line list as in \cite{2020dorazi}. We derived abundances for 11 atomic species, both $\alpha$-, proton-capture and iron-peak elements. Overall, we find a good agreement between optical and NIR abundances. The lack of trends between [X/H] and \teff confirm that our analysis is reliable, with the exception of Cr\,{\sc ii}, for which instead we observed increasing abundances at decreasing temperatures. This trend confirms the previous findings of \cite{2006schuler,2010schuler} about the over-ionisation effects. 
Our derivation of C\,{\sc i} abundances from optical atomic lines reveals a similar effect. The two lines used have high-excitation potential and they yield higher abundances (up to almost +1.0\,dex) at decreasing \teff. In the NIR, we analysed another high-excitation line, 16021\AA\,, but only in two stars, TYC\,1991-1235-1 and HIP\,61205. Despite what we obtained from the optical lines, in the NIR we do not see the same effect as in the optical. \cite{2015schuler} find a trend similar to what we observed for two C lines with $\chi$>7eV and these authors suggested that the C abundance from C$_2$ features is more reliable. In a similar way, we derived C abundances from CH molecular band at 4300\AA\,. At variance with what we obtained from the atomic lines, we did not observe the same trend for the new abundance determinations, in agreement with the findings of \cite{2015schuler}. We suggest that for very young and cool stars the C abundance derived from molecular lines is more reliable.  
The over-ionisation/excitation effects could be explained by a combination of different factors, such as the higher level of activity due to the young age of the stars and the presence of intense local chromospheric and/or photospheric magnetic fields that can alter the line profiles. Indeed, we find a positive correlation between the C abundances derived from the atomic lines in the optical range and the activity indexes $\log$R$^{\prime}_{\rm{HK}}$, suggesting that these effects are related to higher activity levels. This behaviour is not seen in the C estimates from CH molecular features. We also find a positive correlation between the Cr\,{\sc ii} values and $\log$R$^{\prime}_{\rm{HK}}$. However, as already pointed out in previous studies \citep{2020baratella,2020spina}, the main causes are still unknown and they may be a combination of different factors, most likely a combination of more intense chromospheric or photospheric magnetic fields. Finding a theoretical explanation to these issues is beyond the scopes of this paper, but the topic is interesting and deserves a deeper investigation.

\begin{acknowledgements}
We thank the anonymous referee for her/his very helpful comments and suggestions.
The authors acknowledge support by INAF/WOW and INAF/FRONTIERA through the "Progetti Premiali" funding scheme of the Italian Ministry of Education, University, and Research.
\end{acknowledgements}

\bibliographystyle{aa} 
\bibliography{ref} 

\begin{thebibliography}{107}
\expandafter\ifx\csname natexlab\endcsname\relax\def\natexlab#1{#1}\fi

\bibitem[{{Adibekyan}(2019)}]{2019adibekyan}
{Adibekyan}, V. 2019, Geosciences, 9, 105

\bibitem[{{Aleo} {et~al.}(2017){Aleo}, {Sobotka}, \& {Ram{\'\i}rez}}]{2017aleo}
{Aleo}, P.~D., {Sobotka}, A.~C., \& {Ram{\'\i}rez}, I. 2017, \apj, 846, 24

\bibitem[{{Amarsi} {et~al.}(2019){Amarsi}, {Nissen}, \&
  {Sk{\'u}lad{\'o}ttir}}]{2019amarsi}
{Amarsi}, A.~M., {Nissen}, P.~E., \& {Sk{\'u}lad{\'o}ttir}, {\'A}. 2019, \aap,
  630, A104

\bibitem[{{Ammler-von Eiff} \& {Guenther}(2009)}]{2009ammler}
{Ammler-von Eiff}, M. \& {Guenther}, E.~W. 2009, \aap, 508, 677

\bibitem[{{Asplund} {et~al.}(2009){Asplund}, {Grevesse}, {Sauval}, \&
  {Scott}}]{asplund}
{Asplund}, M., {Grevesse}, N., {Sauval}, A.~J., \& {Scott}, P. 2009, \araa, 47,
  481

\bibitem[{{Bailer-Jones} {et~al.}(2018){Bailer-Jones}, {Rybizki}, {Fouesneau},
  {Mantelet}, \& {Andrae}}]{2018bailer}
{Bailer-Jones}, C.~A.~L., {Rybizki}, J., {Fouesneau}, M., {Mantelet}, G., \&
  {Andrae}, R. 2018, \aj, 156, 58

\bibitem[{{Baratella} {et~al.}(2020){Baratella}, {D'Orazi}, {Carraro},
  {Desidera}, {Randich}, {Magrini}, {Adibekyan}, {Smiljanic}, {Spina},
  {Tsantaki}, {Tautvai{\v{s}}ien{\.{e}}}, {Sousa}, {Jofr{\'e}},
  {Jim{\'e}nez-Esteban}, {Delgado-Mena}, {Martell}, {Van der Swaelmen},
  {Roccatagliata}, {Gilmore}, {Alfaro}, {Bayo}, {Bensby}, {Bragaglia},
  {Franciosini}, {Gonneau}, {Heiter}, {Hourihane}, {Jeffries}, {Koposov},
  {Morbidelli}, {Prisinzano}, {Sacco}, {Sbordone}, {Worley}, {Zaggia}, \&
  {Lewis}}]{2020baratella}
{Baratella}, M., {D'Orazi}, V., {Carraro}, G., {et~al.} 2020, \aap, 634, A34

\bibitem[{{Barklem} {et~al.}(2000){Barklem}, {Piskunov}, \&
  {O'Mara}}]{2000barklem}
{Barklem}, P.~S., {Piskunov}, N., \& {O'Mara}, B.~J. 2000, \aaps, 142, 467

\bibitem[{{Bensby} {et~al.}(2014){Bensby}, {Feltzing}, \& {Oey}}]{2014bensby}
{Bensby}, T., {Feltzing}, S., \& {Oey}, M.~S. 2014, \aap, 562, A71

\bibitem[{{Biazzo} {et~al.}(2012){Biazzo}, {D'Orazi}, {Desidera}, {Covino},
  {Alcal{\'a}}, \& {Zusi}}]{2012biazzo}
{Biazzo}, K., {D'Orazi}, V., {Desidera}, S., {et~al.} 2012, \mnras, 427, 2905

\bibitem[{{Biazzo} {et~al.}(2011{\natexlab{a}}){Biazzo}, {Randich}, \&
  {Palla}}]{2011biazzoA}
{Biazzo}, K., {Randich}, S., \& {Palla}, F. 2011{\natexlab{a}}, \aap, 525, A35

\bibitem[{{Biazzo} {et~al.}(2011{\natexlab{b}}){Biazzo}, {Randich}, {Palla}, \&
  {Brice{\~n}o}}]{2011biazzoB}
{Biazzo}, K., {Randich}, S., {Palla}, F., \& {Brice{\~n}o}, C.
  2011{\natexlab{b}}, \aap, 530, A19

\bibitem[{{Blanco-Cuaresma} {et~al.}(2015){Blanco-Cuaresma}, {Soubiran},
  {Heiter}, {Asplund}, {Carraro}, {Costado}, {Feltzing},
  {Gonz{\'a}lez-Hern{\'a}ndez}, {Jim{\'e}nez-Esteban}, {Korn}, {Marino},
  {Montes}, {San Roman}, {Tabernero}, \&
  {Tautvai{\v{s}}ien{\.{e}}}}]{2015blanco-cuaresma}
{Blanco-Cuaresma}, S., {Soubiran}, C., {Heiter}, U., {et~al.} 2015, \aap, 577,
  A47

\bibitem[{{Brewer} {et~al.}(2017){Brewer}, {Fischer}, \&
  {Madhusudhan}}]{2017brewer}
{Brewer}, J.~M., {Fischer}, D.~A., \& {Madhusudhan}, N. 2017, \aj, 153, 83

\bibitem[{{Brewer} {et~al.}(2016){Brewer}, {Fischer}, {Valenti}, \&
  {Piskunov}}]{2016brewer}
{Brewer}, J.~M., {Fischer}, D.~A., {Valenti}, J.~A., \& {Piskunov}, N. 2016,
  \apjs, 225, 32

\bibitem[{{Burkhart} \& {Coupry}(2000)}]{2000burkhart}
{Burkhart}, C. \& {Coupry}, M.~F. 2000, \aap, 354, 216

\bibitem[{{Caffau} {et~al.}(2019){Caffau}, {Bonifacio}, {Oliva}, {Korotin},
  {Capitanio}, {Andrievsky}, {Collet}, {Sbordone}, {Duffau}, {Sanna}, {Tozzi},
  {Origlia}, {Ryde}, \& {Ludwig}}]{2019caffau}
{Caffau}, E., {Bonifacio}, P., {Oliva}, E., {et~al.} 2019, \aap, 622, A68

\bibitem[{{Carleo} {et~al.}(2018){Carleo}, {Benatti}, {Lanza}, {Gratton},
  {Claudi}, {Desidera}, {Mace}, {Messina}, {Sanna}, {Sissa}, {Ghedina},
  {Ghinassi}, {Guerra}, {Harutyunyan}, {Micela}, {Molinari}, {Oliva}, {Tozzi},
  {Baffa}, {Baruffolo}, {Bignamini}, {Buchschacher}, {Cecconi}, {Cosentino},
  {Endl}, {Falcini}, {Fantinel}, {Fini}, {Fugazza}, {Galli}, {Giani},
  {Gonz{\'a}lez}, {Gonz{\'a}lez-{\'A}lvarez}, {Gonz{\'a}lez}, {Hernandez},
  {Hernandez Diaz}, {Iuzzolino}, {Kaplan}, {Kidder}, {Lodi}, {Malavolta},
  {Maldonado}, {Origlia}, {Perez Ventura}, {Puglisi}, {Rainer}, {Riverol},
  {Riverol}, {San Juan}, {Scuderi}, {Seemann}, {Sokal}, {Sozzetti}, \&
  {Sozzi}}]{2018carleo}
{Carleo}, I., {Benatti}, S., {Lanza}, A.~F., {et~al.} 2018, \aap, 613, A50

\bibitem[{{Carleo} {et~al.}(2020){Carleo}, {Malavolta}, {Lanza}, {Damasso},
  {Desidera}, {Borsa}, {Mallonn}, {Pinamonti}, {Gratton}, {Alei}, {Benatti},
  {Mancini}, {Maldonado}, {Biazzo}, {Esposito}, {Frustagli},
  {Gonz{\'a}lez-{\'A}lvarez}, {Micela}, {Scandariato}, {Sozzetti}, {Affer},
  {Bignamini}, {Bonomo}, {Claudi}, {Cosentino}, {Covino}, {Fiorenzano},
  {Giacobbe}, {Harutyunyan}, {Leto}, {Maggio}, {Molinari}, {Nascimbeni},
  {Pagano}, {Pedani}, {Piotto}, {Poretti}, {Rainer}, {Redfield}, {Baffa},
  {Baruffolo}, {Buchschacher}, {Billotti}, {Cecconi}, {Falcini}, {Fantinel},
  {Fini}, {Galli}, {Ghedina}, {Ghinassi}, {Giani}, {Gonzalez}, {Gonzalez},
  {Guerra}, {Hernandez Diaz}, {Hernandez}, {Iuzzolino}, {Lodi}, {Oliva},
  {Origlia}, {Perez Ventura}, {Puglisi}, {Riverol}, {Riverol}, {San Juan},
  {Sanna}, {Scuderi}, {Seemann}, {Sozzi}, \& {Tozzi}}]{2020carleo}
{Carleo}, I., {Malavolta}, L., {Lanza}, A.~F., {et~al.} 2020, arXiv e-prints,
  arXiv:2002.10562

\bibitem[{{Carleo} {et~al.}(2016){Carleo}, {Sanna}, {Gratton}, {Benatti},
  {Bonavita}, {Oliva}, {Origlia}, {Desidera}, {Claudi}, \&
  {Sissa}}]{2016carleo}
{Carleo}, I., {Sanna}, N., {Gratton}, R., {et~al.} 2016, Experimental
  Astronomy, 41, 351

\bibitem[{{Casagrande} {et~al.}(2010){Casagrande}, {Ram{\'\i}rez},
  {Mel{\'e}ndez}, {Bessell}, \& {Asplund}}]{casagrande}
{Casagrande}, L., {Ram{\'\i}rez}, I., {Mel{\'e}ndez}, J., {Bessell}, M., \&
  {Asplund}, M. 2010, \aap, 512, A54

\bibitem[{{Castelli} \& {Kurucz}(2003)}]{2003cast}
{Castelli}, F. \& {Kurucz}, R.~L. 2003, in IAU Symposium, Vol. 210, Modelling
  of Stellar Atmospheres, ed. N.~{Piskunov}, W.~W. {Weiss}, \& D.~F. {Gray},
  A20

\bibitem[{{Claudi} {et~al.}(2017){Claudi}, {Benatti}, {Carleo}, {Ghedina},
  {Guerra}, {Micela}, {Molinari}, {Oliva}, {Rainer}, {Tozzi}, {Baffa},
  {Baruffolo}, {Buchschacher}, {Cecconi}, {Cosentino}, {Fantinel}, {Fini},
  {Ghinassi}, {Giani}, {Gonzalez}, {Gonzalez}, {Gratton}, {Harutyunyan},
  {Hernandez}, {Lodi}, {Malavolta}, {Maldonado}, {Origlia}, {Sanna}, {Sanjuan},
  {Scuderi}, {Seemann}, {Sozzetti}, {Perez Ventura}, {Hernandez Diaz}, {Galli},
  {Gonzalez}, {Riverol}, \& {Riverol}}]{2017claudi}
{Claudi}, R., {Benatti}, S., {Carleo}, I., {et~al.} 2017, European Physical
  Journal Plus, 132, 364

\bibitem[{{Cosentino} {et~al.}(2014){Cosentino}, {Lovis}, {Pepe}, {Cameron},
  {Latham}, {Molinari}, {Udry}, {Bezawada}, {Buchschacher}, {Figueira},
  {Fleury}, {Ghedina}, {Glenday}, {Gonzalez}, {Guerra}, {Henry}, {Hughes},
  {Maire}, {Motalebi}, \& {Phillips}}]{2014cosentino}
{Cosentino}, R., {Lovis}, C., {Pepe}, F., {et~al.} 2014, Society of
  Photo-Optical Instrumentation Engineers (SPIE) Conference Series, Vol. 9147,
  {HARPS-N @ TNG, two year harvesting data: performances and results}, 91478C

\bibitem[{{Covino} {et~al.}(2013){Covino}, {Esposito}, {Barbieri}, {Mancini},
  {Nascimbeni}, {Claudi}, {Desidera}, {Gratton}, {Lanza}, {Sozzetti}, {Biazzo},
  {Affer}, {Gandolfi}, {Munari}, {Pagano}, {Bonomo}, {Collier Cameron},
  {H{\'e}brard}, {Maggio}, {Messina}, {Micela}, {Molinari}, {Pepe}, {Piotto},
  {Ribas}, {Santos}, {Southworth}, {Shkolnik}, {Triaud}, {Bedin}, {Benatti},
  {Boccato}, {Bonavita}, {Borsa}, {Borsato}, {Brown}, {Carolo}, {Ciceri},
  {Cosentino}, {Damasso}, {Faedi}, {Mart{\'\i}nez Fiorenzano}, {Latham},
  {Lovis}, {Mordasini}, {Nikolov}, {Poretti}, {Rainer}, {Rebolo L{\'o}pez},
  {Scandariato}, {Silvotti}, {Smareglia}, {Alcal{\'a}}, {Cunial}, {Di
  Fabrizio}, {Di Mauro}, {Giacobbe}, {Granata}, {Harutyunyan}, {Knapic},
  {Lattanzi}, {Leto}, {Lodato}, {Malavolta}, {Marzari}, {Molinaro},
  {Nardiello}, {Pedani}, {Prisinzano}, \& {Turrini}}]{2013covino}
{Covino}, E., {Esposito}, M., {Barbieri}, M., {et~al.} 2013, \aap, 554, A28

\bibitem[{{Cutri} {et~al.}(2003){Cutri}, {Skrutskie}, {van Dyk}, {Beichman},
  {Carpenter}, {Chester}, {Cambresy}, {Evans}, {Fowler}, {Gizis}, {Howard},
  {Huchra}, {Jarrett}, {Kopan}, {Kirkpatrick}, {Light}, {Marsh}, {McCallon},
  {Schneider}, {Stiening}, {Sykes}, {Weinberg}, {Wheaton}, {Wheelock}, \&
  {Zacarias}}]{2003cutri}
{Cutri}, R.~M., {Skrutskie}, M.~F., {van Dyk}, S., {et~al.} 2003, {2MASS All
  Sky Catalog of point sources.}

\bibitem[{{De Silva} {et~al.}(2015){De Silva}, {Freeman}, {Bland-Hawthorn},
  {Martell}, {de Boer}, {Asplund}, {Keller}, {Sharma}, {Zucker}, {Zwitter},
  {Anguiano}, {Bacigalupo}, {Bayliss}, {Beavis}, {Bergemann}, {Campbell},
  {Cannon}, {Carollo}, {Casagrande}, {Casey}, {Da Costa}, {D'Orazi}, {Dotter},
  {Duong}, {Heger}, {Ireland}, {Kafle}, {Kos}, {Lattanzio}, {Lewis}, {Lin},
  {Lind}, {Munari}, {Nataf}, {O'Toole}, {Parker}, {Reid}, {Schlesinger},
  {Sheinis}, {Simpson}, {Stello}, {Ting}, {Traven}, {Watson}, {Wittenmyer},
  {Yong}, \& {{\v{Z}}erjal}}]{2015galah}
{De Silva}, G.~M., {Freeman}, K.~C., {Bland-Hawthorn}, J., {et~al.} 2015,
  \mnras, 449, 2604

\bibitem[{{D'Orazi} {et~al.}(2011){D'Orazi}, {Biazzo}, \&
  {Randich}}]{2011dorazi}
{D'Orazi}, V., {Biazzo}, K., \& {Randich}, S. 2011, \aap, 526, A103

\bibitem[{{D'Orazi} {et~al.}(2017){D'Orazi}, {Desidera}, {Gratton}, {Lanza},
  {Messina}, {Andrievsky}, {Korotin}, {Benatti}, {Bonnefoy}, {Covino}, \&
  {Janson}}]{2017dorazi}
{D'Orazi}, V., {Desidera}, S., {Gratton}, R.~G., {et~al.} 2017, \aap, 598, A19

\bibitem[{{D'Orazi} {et~al.}(2020){D'Orazi}, {Oliva}, {Bragaglia}, {Frasca},
  {Sanna}, {Biazzo}, {Casali}, {Desidera}, {Lucatello}, {Magrini}, \&
  {Origlia}}]{2020dorazi}
{D'Orazi}, V., {Oliva}, E., {Bragaglia}, A., {et~al.} 2020, \aap, 633, A38

\bibitem[{{D'Orazi} \& {Randich}(2009)}]{2009dorazi}
{D'Orazi}, V. \& {Randich}, S. 2009, \aap, 501, 553

\bibitem[{{D'Orazi} {et~al.}(2009){D'Orazi}, {Randich}, {Flaccomio}, {Palla},
  {Sacco}, \& {Pallavicini}}]{2009dorOrion}
{D'Orazi}, V., {Randich}, S., {Flaccomio}, E., {et~al.} 2009, \aap, 501, 973

\bibitem[{{Dutra-Ferreira} {et~al.}(2016){Dutra-Ferreira}, {Pasquini},
  {Smiljanic}, {Porto de Mello}, \& {Steffen}}]{ferreira}
{Dutra-Ferreira}, L., {Pasquini}, L., {Smiljanic}, R., {Porto de Mello}, G.~F.,
  \& {Steffen}, M. 2016, \aap, 585, A75

\bibitem[{{Eisenbeiss} {et~al.}(2013){Eisenbeiss}, {Ammler-von Eiff}, {Roell},
  {Mugrauer}, {Adam}, {Neuh{\"a}user}, {Schmidt}, \&
  {Bedalov}}]{2013eisenbeiss}
{Eisenbeiss}, T., {Ammler-von Eiff}, M., {Roell}, T., {et~al.} 2013, \aap, 556,
  A53

\bibitem[{{Folsom} {et~al.}(2016){Folsom}, {Petit}, {Bouvier}, {L{\`e}bre},
  {Amard}, {Palacios}, {Morin}, {Donati}, {Jeffers}, {Marsden}, \&
  {Vidotto}}]{2016folsom}
{Folsom}, C.~P., {Petit}, P., {Bouvier}, J., {et~al.} 2016, \mnras, 457, 580

\bibitem[{{Friel} \& {Boesgaard}(1992)}]{1992friel}
{Friel}, E.~D. \& {Boesgaard}, A.~M. 1992, \apj, 387, 170

\bibitem[{{Gebran} {et~al.}(2008){Gebran}, {Monier}, \& {Richard}}]{2008gebran}
{Gebran}, M., {Monier}, R., \& {Richard}, O. 2008, \aap, 479, 189

\bibitem[{{Ghezzi} {et~al.}(2010){Ghezzi}, {Cunha}, {Smith}, {de Ara{\'u}jo},
  {Schuler}, \& {de la Reza}}]{2010ghezzi}
{Ghezzi}, L., {Cunha}, K., {Smith}, V.~V., {et~al.} 2010, \apj, 720, 1290

\bibitem[{{Gilmore} {et~al.}(2012){Gilmore}, {Randich}, {Asplund}, {Binney},
  {Bonifacio}, {Drew}, {Feltzing}, {Ferguson}, {Jeffries}, {Micela},
  {Negueruela}, {Prusti}, {Rix}, {Vallenari}, {Alfaro}, {Allende-Prieto},
  {Babusiaux}, {Bensby}, {Blomme}, {Bragaglia}, {Flaccomio}, {Fran{\c{c}}ois},
  {Irwin}, {Koposov}, {Korn}, {Lanzafame}, {Pancino}, {Paunzen},
  {Recio-Blanco}, {Sacco}, {Smiljanic}, {Van Eck}, {Walton}, {Aden}, {Aerts},
  {Affer}, {Alcala}, {Altavilla}, {Alves}, {Antoja}, {Arenou}, {Argiroffi},
  {Asensio Ramos}, {Bailer-Jones}, {Balaguer-Nunez}, {Bayo}, {Barbuy},
  {Barisevicius}, {Barrado y Navascues}, {Battistini}, {Bellas Velidis},
  {Bellazzini}, {Belokurov}, {Bergemann}, {Bertelli}, {Biazzo}, {Bienayme},
  {Bland-Hawthorn}, {Boeche}, {Bonito}, {Boudreault}, {Bouvier}, {Brandao},
  {Brown}, {de Bruijne}, {Burleigh}, {Caballero}, {Caffau}, {Calura},
  {Capuzzo-Dolcetta}, {Caramazza}, {Carraro}, {Casagrande}, {Casewell},
  {Chapman}, {Chiappini}, {Chorniy}, {Christlieb}, {Cignoni}, {Cocozza},
  {Colless}, {Collet}, {Collins}, {Correnti}, {Covino}, {Crnojevic}, {Cropper},
  {Cunha}, {Damiani}, {David}, {Delgado}, {Duffau}, {Edvardsson}, {Eldridge},
  {Enke}, {Eriksson}, {Evans}, {Eyer}, {Famaey}, {Fellhauer}, {Ferreras},
  {Figueras}, {Fiorentino}, {Flynn}, {Folha}, {Franciosini}, {Frasca},
  {Freeman}, {Fremat}, {Friel}, {Gaensicke}, {Gameiro}, {Garzon}, {Geier},
  {Geisler}, {Gerhard}, {Gibson}, {Gomboc}, {Gomez}, {Gonzalez-Fernandez},
  {Gonzalez Hernandez}, {Gosset}, {Grebel}, {Greimel}, {Groenewegen},
  {Grundahl}, {Guarcello}, {Gustafsson}, {Hadrava}, {Hatzidimitriou}, {Hambly},
  {Hammersley}, {Hansen}, {Haywood}, {Heber}, {Heiter}, {Held}, {Helmi},
  {Hensler}, {Herrero}, {Hill}, {Hodgkin}, {Huelamo}, {Huxor}, {Ibata},
  {Jackson}, {de Jong}, {Jonker}, {Jordan}, {Jordi}, {Jorissen}, {Katz},
  {Kawata}, {Keller}, {Kharchenko}, {Klement}, {Klutsch}, {Knude}, {Koch},
  {Kochukhov}, {Kontizas}, {Koubsky}, {Lallement}, {de Laverny}, {van Leeuwen},
  {Lemasle}, {Lewis}, {Lind}, {Lindstrom}, {Lobel}, {Lopez Santiago}, {Lucas},
  {Ludwig}, {Lueftinger}, {Magrini}, {Maiz Apellaniz}, {Maldonado}, {Marconi},
  {Marino}, {Martayan}, {Martinez-Valpuesta}, {Matijevic}, {McMahon},
  {Messina}, {Meyer}, {Miglio}, {Mikolaitis}, {Minchev}, {Minniti}, {Moitinho},
  {Momany}, {Monaco}, {Montalto}, {Monteiro}, {Monier}, {Montes}, {Mora},
  {Moraux}, {Morel}, {Mowlavi}, {Mucciarelli}, {Munari}, {Napiwotzki},
  {Nardetto}, {Naylor}, {Naze}, {Nelemans}, {Okamoto}, {Ortolani}, {Pace},
  {Palla}, {Palous}, {Parker}, {Penarrubia}, {Pillitteri}, {Piotto}, {Posbic},
  {Prisinzano}, {Puzeras}, {Quirrenbach}, {Ragaini}, {Read}, {Read}, {Reyle},
  {De Ridder}, {Robichon}, {Robin}, {Roeser}, {Romano}, {Royer}, {Ruchti},
  {Ruzicka}, {Ryan}, {Ryde}, {Santos}, {Sanz Forcada}, {Sarro Baro},
  {Sbordone}, {Schilbach}, {Schmeja}, {Schnurr}, {Schoenrich}, {Scholz},
  {Seabroke}, {Sharma}, {De Silva}, {Smith}, {Solano}, {Sordo}, {Soubiran},
  {Sousa}, {Spagna}, {Steffen}, {Steinmetz}, {Stelzer}, {Stempels},
  {Tabernero}, {Tautvaisiene}, {Thevenin}, {Torra}, {Tosi}, {Tolstoy}, {Turon},
  {Walker}, {Wambsganss}, {Worley}, {Venn}, {Vink}, {Wyse}, {Zaggia},
  {Zeilinger}, {Zoccali}, {Zorec}, {Zucker}, {Zwitter}, \& {Gaia-ESO Survey
  Team}}]{2012gilmore}
{Gilmore}, G., {Randich}, S., {Asplund}, M., {et~al.} 2012, The Messenger, 147,
  25

\bibitem[{{Gonzalez} {et~al.}(2010){Gonzalez}, {Carlson}, \&
  {Tobin}}]{2010gonzalez}
{Gonzalez}, G., {Carlson}, M.~K., \& {Tobin}, R.~W. 2010, \mnras, 403, 1368

\bibitem[{{Harutyunyan} {et~al.}(2018){Harutyunyan}, {Rainer}, {Hernandez},
  {Oliva}, {Guerra}, {Lodi}, {San Juan}, {Bignamini}, {Ghedina}, {Ghinassi},
  {Molinari}, {Benatti}, {Carleo}, {Claudi}, {Micela}, {Tozzi}, {Baffa},
  {Baruffolo}, {Biliotti}, {Buchschacher}, {Cecconi}, {Cosentino}, {Falcini},
  {Fantinel}, {Fini}, {Galli}, {Giani}, {Gonzalez}, {Gonzalez-Alvarez},
  {Gonzalez}, {Gratton}, {Hernandez Diaz}, {Iuzzolino}, {Malavolta},
  {Maldonado}, {Origlia}, {Poretti}, {Perez Ventura}, {Puglisi}, {Riverol},
  {Riverol}, {Sanna}, {Scuderi}, {Seeman}, {Sozzetti}, \&
  {Sozzi}}]{2018harutyunyan}
{Harutyunyan}, A., {Rainer}, M., {Hernandez}, N., {et~al.} 2018, in Society of
  Photo-Optical Instrumentation Engineers (SPIE) Conference Series, Vol. 10706,
  \procspie, 1070642

\bibitem[{Hunter {et~al.}(2012)Hunter, Macgregor, Szabo, Wellington, \&
  Bellgard}]{PMID:22333270}
Hunter, A.~A., Macgregor, A.~B., Szabo, T.~O., Wellington, C.~A., \& Bellgard,
  M.~I. 2012, Source code for biology and medicine, 7, 1

\bibitem[{{James} {et~al.}(2006){James}, {Melo}, {Santos}, \&
  {Bouvier}}]{2006james}
{James}, D.~J., {Melo}, C., {Santos}, N.~C., \& {Bouvier}, J. 2006, \aap, 446,
  971

\bibitem[{{Jofr{\'e}} {et~al.}(2019){Jofr{\'e}}, {Heiter}, \&
  {Soubiran}}]{2019jofre}
{Jofr{\'e}}, P., {Heiter}, U., \& {Soubiran}, C. 2019, \araa, 57, 571

\bibitem[{{King} \& {Schuler}(2005)}]{2005kingschuler}
{King}, J.~R. \& {Schuler}, S.~C. 2005, \pasp, 117, 911

\bibitem[{{King} {et~al.}(2000){King}, {Soderblom}, {Fischer}, \&
  {Jones}}]{2000king}
{King}, J.~R., {Soderblom}, D.~R., {Fischer}, D., \& {Jones}, B.~F. 2000, \apj,
  533, 944

\bibitem[{{Lawler} {et~al.}(2013){Lawler}, {Guzman}, {Wood}, {Sneden}, \&
  {Cowan}}]{lawler13}
{Lawler}, J.~E., {Guzman}, A., {Wood}, M.~P., {Sneden}, C., \& {Cowan}, J.~J.
  2013, \apjs, 205, 11

\bibitem[{{Lawler} {et~al.}(2017){Lawler}, {Sneden}, {Nave}, {Den Hartog},
  {Emraho{\u{g}}lu}, \& {Cowan}}]{2017lawler}
{Lawler}, J.~E., {Sneden}, C., {Nave}, G., {et~al.} 2017, \apjs, 228, 10

\bibitem[{{Lind} {et~al.}(2011){Lind}, {Asplund}, {Barklem}, \&
  {Belyaev}}]{2011lind}
{Lind}, K., {Asplund}, M., {Barklem}, P.~S., \& {Belyaev}, A.~K. 2011, \aap,
  528, A103

\bibitem[{{Lindegren} {et~al.}(2018){Lindegren}, {Hern{\'a}ndez}, {Bombrun},
  {Klioner}, {Bastian}, {Ramos-Lerate}, {de Torres}, {Steidelm{\"u}ller},
  {Stephenson}, {Hobbs}, {Lammers}, {Biermann}, {Geyer}, {Hilger}, {Michalik},
  {Stampa}, {McMillan}, {Casta{\~n}eda}, {Clotet}, {Comoretto}, {Davidson},
  {Fabricius}, {Gracia}, {Hambly}, {Hutton}, {Mora}, {Portell}, {van Leeuwen},
  {Abbas}, {Abreu}, {Altmann}, {Andrei}, {Anglada}, {Balaguer-N{\'u}{\~n}ez},
  {Barache}, {Becciani}, {Bertone}, {Bianchi}, {Bouquillon}, {Bourda},
  {Br{\"u}semeister}, {Bucciarelli}, {Busonero}, {Buzzi}, {Cancelliere},
  {Carlucci}, {Charlot}, {Cheek}, {Crosta}, {Crowley}, {de Bruijne}, {de
  Felice}, {Drimmel}, {Esquej}, {Fienga}, {Fraile}, {Gai}, {Garralda},
  {Gonz{\'a}lez-Vidal}, {Guerra}, {Hauser}, {Hofmann}, {Holl}, {Jordan},
  {Lattanzi}, {Lenhardt}, {Liao}, {Licata}, {Lister}, {L{\"o}ffler},
  {Marchant}, {Martin-Fleitas}, {Messineo}, {Mignard}, {Morbidelli}, {Poggio},
  {Riva}, {Rowell}, {Salguero}, {Sarasso}, {Sciacca}, {Siddiqui}, {Smart},
  {Spagna}, {Steele}, {Taris}, {Torra}, {van Elteren}, {van Reeven}, \&
  {Vecchiato}}]{2018lindegren}
{Lindegren}, L., {Hern{\'a}ndez}, J., {Bombrun}, A., {et~al.} 2018, \aap, 616,
  A2

\bibitem[{{Linsky} {et~al.}(2012){Linsky}, {Bushinsky}, {Ayres}, {Fontenla}, \&
  {France}}]{2012linsky}
{Linsky}, J.~L., {Bushinsky}, R., {Ayres}, T., {Fontenla}, J., \& {France}, K.
  2012, \apj, 745, 25

\bibitem[{{Lodders}(2003)}]{2003lodders}
{Lodders}, K. 2003, \apj, 591, 1220

\bibitem[{{L{\'o}pez-Santiago} {et~al.}(2006){L{\'o}pez-Santiago}, {Montes},
  {Crespo-Chac{\'o}n}, \& {Fern{\'a}ndez-Figueroa}}]{2006lopez}
{L{\'o}pez-Santiago}, J., {Montes}, D., {Crespo-Chac{\'o}n}, I., \&
  {Fern{\'a}ndez-Figueroa}, M.~J. 2006, \apj, 643, 1160

\bibitem[{{Lovis} {et~al.}(2011){Lovis}, {Dumusque}, {Santos}, {Bouchy},
  {Mayor}, {Pepe}, {Queloz}, {S{\'e}gransan}, \& {Udry}}]{2011lovis}
{Lovis}, C., {Dumusque}, X., {Santos}, N.~C., {et~al.} 2011, arXiv e-prints,
  arXiv:1107.5325

\bibitem[{{Luck}(2017)}]{2017luck}
{Luck}, R.~E. 2017, \aj, 153, 21

\bibitem[{{Malavolta} {et~al.}(2016){Malavolta}, {Nascimbeni}, {Piotto},
  {Quinn}, {Borsato}, {Granata}, {Bonomo}, {Marzari}, {Bedin}, {Rainer},
  {Desidera}, {Lanza}, {Poretti}, {Sozzetti}, {White}, {Latham}, {Cunial},
  {Libralato}, {Nardiello}, {Boccato}, {Claudi}, {Cosentino}, {Covino},
  {Gratton}, {Maggio}, {Micela}, {Molinari}, {Pagano}, {Smareglia}, {Affer},
  {Andreuzzi}, {Aparicio}, {Benatti}, {Bignamini}, {Borsa}, {Damasso}, {Di
  Fabrizio}, {Harutyunyan}, {Esposito}, {Fiorenzano}, {Gandolfi}, {Giacobbe},
  {Gonz{\'a}lez Hern{\'a}ndez}, {Maldonado}, {Masiero}, {Molinaro}, {Pedani},
  \& {Scandariato}}]{2016malavolta}
{Malavolta}, L., {Nascimbeni}, V., {Piotto}, G., {et~al.} 2016, \aap, 588, A118

\bibitem[{{Maldonado} {et~al.}(2015){Maldonado}, {Eiroa}, {Villaver},
  {Montesinos}, \& {Mora}}]{2015maldonado}
{Maldonado}, J., {Eiroa}, C., {Villaver}, E., {Montesinos}, B., \& {Mora}, A.
  2015, \aap, 579, A20

\bibitem[{{Marfil} {et~al.}(2020){Marfil}, {Tabernero}, {Montes}, {Caballero},
  {Soto}, {Gonz{\'a}lez Hern{\'a}ndez}, {Kaminski}, {Nagel}, {Jeffers},
  {Reiners}, {Ribas}, {Quirrenbach}, \& {Amado}}]{2020marfil}
{Marfil}, E., {Tabernero}, H.~M., {Montes}, D., {et~al.} 2020, \mnras, 492,
  5470

\bibitem[{{Marsden} {et~al.}(2014){Marsden}, {Petit}, {Jeffers}, {Morin},
  {Fares}, {Reiners}, {do Nascimento}, {Auri{\`e}re}, {Bouvier}, {Carter},
  {Catala}, {Dintrans}, {Donati}, {Gastine}, {Jardine}, {Konstantinova-Antova},
  {Lanoux}, {Ligni{\`e}res}, {Morgenthaler}, {Ram{\`\i}rez-V{\`e}lez},
  {Th{\'e}ado}, {Van Grootel}, \& {BCool Collaboration}}]{2014marsden}
{Marsden}, S.~C., {Petit}, P., {Jeffers}, S.~V., {et~al.} 2014, \mnras, 444,
  3517

\bibitem[{{Mart{\'\i}nez-Arn{\'a}iz} {et~al.}(2010){Mart{\'\i}nez-Arn{\'a}iz},
  {Maldonado}, {Montes}, {Eiroa}, \& {Montesinos}}]{2010martinez}
{Mart{\'\i}nez-Arn{\'a}iz}, R., {Maldonado}, J., {Montes}, D., {Eiroa}, C., \&
  {Montesinos}, B. 2010, \aap, 520, A79

\bibitem[{{Mel{\'e}ndez}(2004)}]{2004melendez}
{Mel{\'e}ndez}, J. 2004, \apj, 615, 1042

\bibitem[{{Mel{\'e}ndez} {et~al.}(2009){Mel{\'e}ndez}, {Asplund}, {Gustafsson},
  \& {Yong}}]{2009melendez}
{Mel{\'e}ndez}, J., {Asplund}, M., {Gustafsson}, B., \& {Yong}, D. 2009, \apjl,
  704, L66

\bibitem[{{Mel{\'e}ndez} {et~al.}(2014){Mel{\'e}ndez}, {Ram{\'\i}rez},
  {Karakas}, {Yong}, {Monroe}, {Bedell}, {Bergemann}, {Asplund}, {Tucci Maia},
  {Bean}, {do Nascimento}, {Bazot}, {Alves-Brito}, {Freitas}, \&
  {Castro}}]{2014melendez}
{Mel{\'e}ndez}, J., {Ram{\'\i}rez}, I., {Karakas}, A. a.~I., {et~al.} 2014,
  \apj, 791, 14

\bibitem[{{Mermilliod} {et~al.}(2008){Mermilliod}, {Grenon}, \&
  {Mayor}}]{2008mermilliod}
{Mermilliod}, J.~C., {Grenon}, M., \& {Mayor}, M. 2008, \aap, 491, 951

\bibitem[{{Mermilliod} {et~al.}(2009){Mermilliod}, {Mayor}, \&
  {Udry}}]{2009mermilliod}
{Mermilliod}, J.~C., {Mayor}, M., \& {Udry}, S. 2009, \aap, 498, 949

\bibitem[{{Monier}(2005)}]{2005monier}
{Monier}, R. 2005, \aap, 442, 563

\bibitem[{{Montes} {et~al.}(2001){Montes}, {L{\'o}pez-Santiago}, {G{\'a}lvez},
  {Fern{\'a}ndez-Figueroa}, {De Castro}, \& {Cornide}}]{2001montes}
{Montes}, D., {L{\'o}pez-Santiago}, J., {G{\'a}lvez}, M.~C., {et~al.} 2001,
  \mnras, 328, 45

\bibitem[{{Netopil} {et~al.}(2016){Netopil}, {Paunzen}, {Heiter}, \&
  {Soubiran}}]{2016netopil}
{Netopil}, M., {Paunzen}, E., {Heiter}, U., \& {Soubiran}, C. 2016, \aap, 585,
  A150

\bibitem[{{Nissen}(2015)}]{2015nissen}
{Nissen}, P.~E. 2015, \aap, 579, A52

\bibitem[{{Noyes} {et~al.}(1984){Noyes}, {Hartmann}, {Baliunas}, {Duncan}, \&
  {Vaughan}}]{1984noyes}
{Noyes}, R.~W., {Hartmann}, L.~W., {Baliunas}, S.~L., {Duncan}, D.~K., \&
  {Vaughan}, A.~H. 1984, \apj, 279, 763

\bibitem[{{Oliva} {et~al.}(2006){Oliva}, {Origlia}, {Baffa}, {Biliotti},
  {Bruno}, {D'Amato}, {Del Vecchio}, {Falcini}, {Gennari}, {Ghinassi}, {Giani},
  {Gonzalez}, {Leone}, {Lolli}, {Lodi}, {Maiolino}, {Mannucci}, {Marcucci},
  {Mochi}, {Montegriffo}, {Rossetti}, {Scuderi}, \& {Sozzi}}]{2006oliva}
{Oliva}, E., {Origlia}, L., {Baffa}, C., {et~al.} 2006, in Society of
  Photo-Optical Instrumentation Engineers (SPIE) Conference Series, Vol. 6269,
  \procspie, 626919

\bibitem[{{Osorio} {et~al.}(2015){Osorio}, {Barklem}, {Lind}, {Belyaev},
  {Spielfiedel}, {Guitou}, \& {Feautrier}}]{2015osorio}
{Osorio}, Y., {Barklem}, P.~S., {Lind}, K., {et~al.} 2015, \aap, 579, A53

\bibitem[{{Poretti} {et~al.}(2016){Poretti}, {Boccato}, {Claudi}, {Cosentino},
  {Covino}, {Desidera}, {Gratton}, {Lanza}, {Maggio}, {Micela}, {Molinari},
  {Pagano}, {Piotto}, {Smareglia}, {Sozzetti}, \& {GAPS
  Collaboration}}]{2016poretti}
{Poretti}, E., {Boccato}, C., {Claudi}, R., {et~al.} 2016, \memsai, 87, 141

\bibitem[{{Rainer} {et~al.}(2018){Rainer}, {Harutyunyan}, {Carleo}, {Oliva},
  {Benatti}, {Bignamini}, {Claudi}, {Gonzalez-Alvarez}, {Sanna}, {Ghedina},
  {Micela}, {Molinari}, {Tozzi}, {Baffa}, {Baruffolo}, {Buchschacher},
  {Cecconi}, {Cosentino}, {Falcini}, {Fantinel}, {Fini}, {Galli}, {Ghinassi},
  {Giani}, {Gonzalez}, {Gonzalez}, {Gratton}, {Guerra}, {Hernandez Diaz},
  {Hernandez}, {Iuzzolino}, {Lodi}, {Malavolta}, {Maldonado}, {Origlia}, {Perez
  Ventura}, {Puglisi}, {Riverol}, {Riverol}, {San Juan}, {Scuderi}, {Seeman},
  {Sozzetti}, \& {Sozzi}}]{2018gofio}
{Rainer}, M., {Harutyunyan}, A., {Carleo}, I., {et~al.} 2018, in Society of
  Photo-Optical Instrumentation Engineers (SPIE) Conference Series, Vol. 10702,
  \procspie, 1070266

\bibitem[{{Ram{\'\i}rez} {et~al.}(2007){Ram{\'\i}rez}, {Allende Prieto}, \&
  {Lambert}}]{2007ramirez}
{Ram{\'\i}rez}, I., {Allende Prieto}, C., \& {Lambert}, D.~L. 2007, Astronomy
  and Astrophysics, 465, 271

\bibitem[{{Ram{\'\i}rez} {et~al.}(2013){Ram{\'\i}rez}, {Allende Prieto}, \&
  {Lambert}}]{2013ramirez}
{Ram{\'\i}rez}, I., {Allende Prieto}, C., \& {Lambert}, D.~L. 2013, \apj, 764,
  78

\bibitem[{{Reiners} {et~al.}(2013){Reiners}, {Shulyak}, {Anglada-Escud{\'e}},
  {Jeffers}, {Morin}, {Zechmeister}, {Kochukhov}, \& {Piskunov}}]{2013reiners}
{Reiners}, A., {Shulyak}, D., {Anglada-Escud{\'e}}, G., {et~al.} 2013, \aap,
  552, A103

\bibitem[{{Ruffoni} {et~al.}(2014){Ruffoni}, {Den Hartog}, {Lawler}, {Brewer},
  {Lind}, {Nave}, \& {Pickering}}]{ruffoni2014}
{Ruffoni}, M.~P., {Den Hartog}, E.~A., {Lawler}, J.~E., {et~al.} 2014, \mnras,
  441, 3127

\bibitem[{{Santos} {et~al.}(2008){Santos}, {Melo}, {James}, {Gameiro},
  {Bouvier}, \& {Gomes}}]{2008santos}
{Santos}, N.~C., {Melo}, C., {James}, D.~J., {et~al.} 2008, \aap, 480, 889

\bibitem[{{Schuler} {et~al.}(2006){Schuler}, {Hatzes}, {King}, {K{\"u}rster},
  \& {The}}]{2006schuler}
{Schuler}, S.~C., {Hatzes}, A.~P., {King}, J.~R., {K{\"u}rster}, M., \& {The},
  L.-S. 2006, \aj, 131, 1057

\bibitem[{{Schuler} {et~al.}(2004){Schuler}, {King}, {Hobbs}, \&
  {Pinsonneault}}]{2004schuler}
{Schuler}, S.~C., {King}, J.~R., {Hobbs}, L.~M., \& {Pinsonneault}, M.~H. 2004,
  \apjl, 602, L117

\bibitem[{{Schuler} {et~al.}(2010){Schuler}, {Plunkett}, {King}, \&
  {Pinsonneault}}]{2010schuler}
{Schuler}, S.~C., {Plunkett}, A.~L., {King}, J.~R., \& {Pinsonneault}, M.~H.
  2010, \pasp, 122, 766

\bibitem[{{Schuler} {et~al.}(2015){Schuler}, {Vaz}, {Katime Santrich}, {Cunha},
  {Smith}, {King}, {Teske}, {Ghezzi}, {Howell}, \& {Isaacson}}]{2015schuler}
{Schuler}, S.~C., {Vaz}, Z.~A., {Katime Santrich}, O. o.~J., {et~al.} 2015,
  \apj, 815, 5

\bibitem[{{Shchukina} {et~al.}(2015){Shchukina}, {Sukhorukov}, \& {Trujillo
  Bueno}}]{2015shchukina}
{Shchukina}, N.~G., {Sukhorukov}, A.~V., \& {Trujillo Bueno}, J. 2015, in IAU
  Symposium, Vol. 305, Polarimetry, ed. K.~N. {Nagendra}, S.~{Bagnulo},
  R.~{Centeno}, \& M.~{Jes{\'u}s Mart{\'\i}nez Gonz{\'a}lez}, 368--371

\bibitem[{{Shetrone} {et~al.}(2015){Shetrone}, {Bizyaev}, {Lawler}, {Allende
  Prieto}, {Johnson}, {Smith}, {Cunha}, {Holtzman}, {Garc{\'\i}a P{\'e}rez},
  {M{\'e}sz{\'a}ros}, {Sobeck}, {Zamora}, {Garc{\'\i}a-Hern{\'a}ndez}, {Souto},
  {Chojnowski}, {Koesterke}, {Majewski}, \& {Zasowski}}]{2015shetrone}
{Shetrone}, M., {Bizyaev}, D., {Lawler}, J.~E., {et~al.} 2015, \apjs, 221, 24

\bibitem[{{Silaj} \& {Landstreet}(2014)}]{2014silaj}
{Silaj}, J. \& {Landstreet}, J.~D. 2014, \aap, 566, A132

\bibitem[{{Sneden} \& {Lambert}(1982)}]{1982sneden}
{Sneden}, C. \& {Lambert}, D.~L. 1982, \apj, 259, 381

\bibitem[{{Sneden} {et~al.}(2014){Sneden}, {Lucatello}, {Ram}, {Brooke}, \&
  {Bernath}}]{2014sneden}
{Sneden}, C., {Lucatello}, S., {Ram}, R.~S., {Brooke}, J. S.~A., \& {Bernath},
  P. 2014, \apjs, 214, 26

\bibitem[{{Sneden}(1973)}]{sneden73}
{Sneden}, C.~A. 1973, PhD thesis, THE UNIVERSITY OF TEXAS AT AUSTIN.

\bibitem[{{Sobeck} {et~al.}(2011){Sobeck}, {Kraft}, {Sneden}, {Preston},
  {Cowan}, {Smith}, {Thompson}, {Shectman}, \& {Burley}}]{2011sob}
{Sobeck}, J.~S., {Kraft}, R.~P., {Sneden}, C., {et~al.} 2011, \aj, 141, 175

\bibitem[{{Soderblom} \& {Mayor}(1993)}]{1993soderblom}
{Soderblom}, D.~R. \& {Mayor}, M. 1993, \aj, 105, 226

\bibitem[{{Sousa} {et~al.}(2007){Sousa}, {Santos}, {Israelian}, {Mayor}, \&
  {Monteiro}}]{sousa}
{Sousa}, S.~G., {Santos}, N.~C., {Israelian}, G., {Mayor}, M., \& {Monteiro},
  M.~J.~P.~F.~G. 2007, \aap, 469, 783

\bibitem[{{Souto} {et~al.}(2018){Souto}, {Cunha}, {Smith}, {Allende Prieto},
  {Garc{\'\i}a-Hern{\'a}ndez}, {Pinsonneault}, {Holzer}, {Frinchaboy},
  {Holtzman}, {Johnson}, {J{\"o}nsson}, {Majewski}, {Shetrone}, {Sobeck},
  {Stringfellow}, {Teske}, {Zamora}, {Zasowski}, {Carrera}, {Stassun}, {Fernand
  ez-Trincado}, {Villanova}, {Minniti}, \& {Santana}}]{2018souto}
{Souto}, D., {Cunha}, K., {Smith}, V.~V., {et~al.} 2018, \apj, 857, 14

\bibitem[{{Sozzetti} {et~al.}(2007){Sozzetti}, {Torres}, {Charbonneau},
  {Latham}, {Holman}, {Winn}, {Laird}, \& {O'Donovan}}]{2007sozzetti}
{Sozzetti}, A., {Torres}, G., {Charbonneau}, D., {et~al.} 2007, \apj, 664, 1190

\bibitem[{{Spina} {et~al.}(2020){Spina}, {Nordlander}, {Casey}, {Bedell},
  {D'Orazi}, {Mel{\'e}ndez}, {Karakas}, {Desidera}, {Baratella}, {Galarza}, \&
  {Casali}}]{2020spina}
{Spina}, L., {Nordlander}, T., {Casey}, A.~R., {et~al.} 2020, arXiv e-prints,
  arXiv:2004.09771

\bibitem[{{Spina} {et~al.}(2017){Spina}, {Randich}, {Magrini}, {Jeffries},
  {Friel}, {Sacco}, {Pancino}, {Bonito}, {Bravi}, {Franciosini}, {Klutsch},
  {Montes}, {Gilmore}, {Vallenari}, {Bensby}, {Bragaglia}, {Flaccomio},
  {Koposov}, {Korn}, {Lanzafame}, {Smiljanic}, {Bayo}, {Carraro}, {Casey},
  {Costado}, {Damiani}, {Donati}, {Frasca}, {Hourihane}, {Jofr{\'e}}, {Lewis},
  {Lind}, {Monaco}, {Morbidelli}, {Prisinzano}, {Sousa}, {Worley}, \&
  {Zaggia}}]{Spina17}
{Spina}, L., {Randich}, S., {Magrini}, L., {et~al.} 2017, \aap, 601, A70

\bibitem[{{Steffen} {et~al.}(2013){Steffen}, {Caffau}, \&
  {Ludwig}}]{2013steffen}
{Steffen}, M., {Caffau}, E., \& {Ludwig}, H.~G. 2013, Memorie della Societa
  Astronomica Italiana Supplementi, 24, 37

\bibitem[{{Tabernero} {et~al.}(2019){Tabernero}, {Marfil}, {Montes}, \&
  {Gonz{\'a}lez Hern{\'a}ndez}}]{2019tabernero}
{Tabernero}, H.~M., {Marfil}, E., {Montes}, D., \& {Gonz{\'a}lez
  Hern{\'a}ndez}, J.~I. 2019, \aap, 628, A131

\bibitem[{{Tabernero} {et~al.}(2017){Tabernero}, {Montes}, {Gonz{\'a}lez
  Hern{\'a}ndez}, \& {Ammler-von Eiff}}]{2017tabernero}
{Tabernero}, H.~M., {Montes}, D., {Gonz{\'a}lez Hern{\'a}ndez}, J.~I., \&
  {Ammler-von Eiff}, M. 2017, \aap, 597, A33

\bibitem[{{Takeda} \& {Honda}(2020)}]{2020takeda}
{Takeda}, Y. \& {Honda}, S. 2020, \aj, 159, 174

\bibitem[{{Teske} {et~al.}(2013){Teske}, {Cunha}, {Schuler}, {Griffith}, \&
  {Smith}}]{2013teske}
{Teske}, J.~K., {Cunha}, K., {Schuler}, S.~C., {Griffith}, C.~A., \& {Smith},
  V.~V. 2013, \apj, 778, 132

\bibitem[{{Tsantaki} {et~al.}(2019){Tsantaki}, {Santos}, {Sousa},
  {Delgado-Mena}, {Adibekyan}, \& {Andreasen}}]{2019tsantaki}
{Tsantaki}, M., {Santos}, N.~C., {Sousa}, S.~G., {et~al.} 2019, \mnras, 485,
  2772

\bibitem[{{Tsantaki} {et~al.}(2013){Tsantaki}, {Sousa}, {Adibekyan}, {Santos},
  {Mortier}, \& {Israelian}}]{2013tsantaki}
{Tsantaki}, M., {Sousa}, S.~G., {Adibekyan}, V.~Z., {et~al.} 2013, \aap, 555,
  A150

\bibitem[{{Valenti} \& {Fischer}(2005)}]{2005valenti}
{Valenti}, J.~A. \& {Fischer}, D.~A. 2005, \apjs, 159, 141

\bibitem[{{Viana Almeida} {et~al.}(2009){Viana Almeida}, {Santos}, {Melo},
  {Ammler-von Eiff}, {Torres}, {Quast}, {Gameiro}, \& {Sterzik}}]{2009viana}
{Viana Almeida}, P., {Santos}, N.~C., {Melo}, C., {et~al.} 2009, \aap, 501, 965

\bibitem[{{Yana Galarza} {et~al.}(2019){Yana Galarza}, {Mel{\'e}ndez},
  {Lorenzo-Oliveira}, {Valio}, {Reggiani}, {Carlos}, {Ponte}, {Spina},
  {Haywood}, \& {Gandolfi}}]{2019galarza}
{Yana Galarza}, J., {Mel{\'e}ndez}, J., {Lorenzo-Oliveira}, D., {et~al.} 2019,
  \mnras, 490, L86

\bibitem[{{Zhao} {et~al.}(2016){Zhao}, {Mashonkina}, {Yan}, {Alexeeva},
  {Kobayashi}, {Pakhomov}, {Shi}, {Sitnova}, {Tan}, {Zhang}, {Zhang}, {Zhou},
  {Bolte}, {Chen}, {Li}, {Liu}, \& {Zhai}}]{2016zhao}
{Zhao}, G., {Mashonkina}, L., {Yan}, H.~L., {et~al.} 2016, \apj, 833, 225

\end{thebibliography}

\clearpage
\onecolumn

\begin{appendix}
\section{Optical line list}
The line list used in the analysis of the HARPS-N optical spectra is shown in Table\,\ref{optical_lines}. The source of oscillator strengths include the NIST database, \cite{lawler13} for Ti lines, line lists published by \cite{2017dorazi}, and \cite{ruffoni2014} for Fe~{\sc i}.

\begin{center}
\begin{longtable}{lccr}
\caption{Line list for the HARPS-N spectra.}
\label{optical_lines}\\ %
\toprule
\multicolumn{1}{l}{Wavelength (\AA)} & \multicolumn{1}{c}{Ion} & \multicolumn{1}{c}{E.P. (eV)} & \multicolumn{1}{r}{$\log gf$}\\ 
\midrule 
\endfirsthead %
 \caption{Continued.}\\ %
  \toprule
\multicolumn{1}{l}{Wavelength (\AA)} & \multicolumn{1}{c}{Ion} & \multicolumn{1}{c}{E.P. (eV)} & \multicolumn{1}{r}{$\log gf$}\\ 
  \midrule  
\endhead %
\bottomrule
\endfoot %
\bottomrule
\endlastfoot %
5380.337 & 6.0 & 7.68 & $-$1.62 \\
6587.610 & 6.0 & 8.54 & $-$1.00 \\
6154.230 & 11.0 & 2.10 & $-$1.57 \\
6160.747 & 11.0 & 2.10 & $-$1.25 \\
4730.029 & 12.0 & 4.35 & $-$2.30 \\
5711.090 & 12.0 & 4.35 & $-$1.71 \\
6318.720 & 12.0 & 5.11 & $-$2.10 \\
6319.240 & 12.0 & 5.11 & $-$2.32 \\
6696.020 & 13.0 & 3.14 & $-$1.62 \\
6698.670 & 13.0 & 3.14 & $-$1.92 \\
5645.610 & 14.0 & 4.93 & $-$2.04 \\
5665.560 & 14.0 & 4.92 & $-$1.94 \\
5684.480 & 14.0 & 4.95 & $-$1.55 \\
5690.425 & 14.0 & 4.93 & $-$1.74 \\
6125.020 & 14.0 & 5.61 & $-$1.52 \\
6142.480 & 14.0 & 5.62 & $-$1.50 \\
6155.130 & 14.0 & 5.62 & $-$0.72 \\
6237.320 & 14.0 & 5.61 & $-$1.05 \\
6243.810 & 14.0 & 5.62 & $-$1.29 \\
6244.470 & 14.0 & 5.62 & $-$1.32 \\
6721.848 & 14.0 & 5.86 & $-$1.13 \\
5260.390 & 20.0 & 2.52 & $-$1.78 \\
5261.700 & 20.0 & 2.52 & $-$0.58 \\
5581.960 & 20.0 & 2.52 & $-$0.67 \\
5857.451 & 20.0 & 2.93 & 0.26 \\
5867.560 & 20.0 & 2.93 & $-$1.60 \\
6169.560 & 20.0 & 2.53 & $-$0.52 \\
6455.600 & 20.0 & 2.52 & $-$1.35 \\
6499.650 & 20.0 & 2.52 & $-$0.81 \\
6508.850 & 20.0 & 2.53 & $-$2.53 \\
4186.120 & 22.0 & 1.50 & $-$0.24 \\
4287.400 & 22.0 & 0.83 & $-$0.37 \\
4427.100 & 22.0 & 1.50 & 0.23 \\
4453.310 & 22.0 & 1.42 & $-$0.03 \\
4453.700 & 22.0 & 1.87 & 0.10 \\
4471.240 & 22.0 & 1.73 & $-$0.15 \\
4518.020 & 22.0 & 0.82 & $-$0.25 \\
4548.760 & 22.0 & 0.82 & $-$0.28 \\
4623.100 & 22.0 & 1.73 & 0.16 \\
4639.660 & 22.0 & 1.74 & $-$0.14 \\
4722.610 & 22.0 & 1.05 & $-$1.47 \\
4758.900 & 22.0 & 0.83 & $-$2.17 \\
4778.250 & 22.0 & 2.23 & $-$0.35 \\
4781.710 & 22.0 & 0.85 & $-$1.95 \\
4797.980 & 22.0 & 2.33 & $-$0.63 \\
4805.410 & 22.0 & 2.34 & 0.07 \\
4820.410 & 22.0 & 1.50 & $-$0.38 \\
4840.870 & 22.0 & 0.90 & $-$0.43 \\
4856.010 & 22.0 & 2.25 & 0.52 \\
4870.120 & 22.0 & 2.24 & 0.44 \\
4885.080 & 22.0 & 1.88 & 0.41 \\
4899.910 & 22.0 & 1.87 & 0.31 \\
4921.760 & 22.0 & 2.17 & 0.04 \\
4937.730 & 22.0 & 0.81 & $-$2.08 \\
4995.070 & 22.0 & 2.24 & $-$1.00 \\
5016.160 & 22.0 & 0.85 & $-$0.48 \\
5020.030 & 22.0 & 0.83 & $-$0.33 \\
5036.460 & 22.0 & 1.44 & 0.14 \\
5038.400 & 22.0 & 1.42 & 0.02 \\
5040.610 & 22.0 & 0.82 & $-$1.67 \\
5043.580 & 22.0 & 0.83 & $-$1.59 \\
5062.100 & 22.0 & 2.16 & $-$0.39 \\
5064.650 & 22.0 & 0.05 & $-$0.94 \\
5087.060 & 22.0 & 1.42 & $-$0.88 \\
5145.460 & 22.0 & 1.46 & $-$0.54 \\
5192.970 & 22.0 & 0.02 & $-$0.95 \\
5210.380 & 22.0 & 0.05 & $-$0.82 \\
5219.700 & 22.0 & 0.02 & $-$2.22 \\
5295.780 & 22.0 & 1.06 & $-$1.59 \\
5389.170 & 22.0 & 0.81 & $-$2.35 \\
5471.190 & 22.0 & 1.44 & $-$1.42 \\
5474.220 & 22.0 & 1.46 & $-$1.23 \\
5503.900 & 22.0 & 2.57 & $-$0.05 \\
5512.520 & 22.0 & 1.46 & $-$0.40 \\
5514.340 & 22.0 & 1.42 & $-$0.66 \\
5514.530 & 22.0 & 1.44 & $-$0.50 \\
5565.470 & 22.0 & 2.23 & $-$0.22 \\
5739.980 & 22.0 & 2.23 & $-$0.92 \\
5785.900 & 22.0 & 3.32 & 0.60 \\
5866.450 & 22.0 & 1.06 & $-$0.79 \\
5880.270 & 22.0 & 1.05 & $-$2.00 \\
5922.110 & 22.0 & 1.04 & $-$1.38 \\
5937.810 & 22.0 & 1.06 & $-$1.94 \\
6091.170 & 22.0 & 2.26 & $-$0.32 \\
6092.790 & 22.0 & 1.88 & $-$1.38 \\
6258.100 & 22.0 & 1.44 & $-$0.39 \\
6261.100 & 22.0 & 1.42 & $-$0.53 \\
6303.760 & 22.0 & 1.44 & $-$1.58 \\
6312.240 & 22.0 & 1.46 & $-$1.55 \\
6554.220 & 22.0 & 1.44 & $-$1.15 \\
6556.060 & 22.0 & 1.46 & $-$1.06 \\
4053.821 & 22.1 & 1.89 & $-$1.07 \\
4163.644 & 22.1 & 2.59 & $-$0.13 \\
4316.794 & 22.1 & 2.05 & $-$1.62 \\
4320.950 & 22.1 & 1.16 & $-$1.88 \\
4395.839 & 22.1 & 1.24 & $-$1.93 \\
4443.801 & 22.1 & 1.08 & $-$0.71 \\
4444.554 & 22.1 & 1.11 & $-$2.20 \\
4468.493 & 22.1 & 1.13 & $-$0.63 \\
4493.522 & 22.1 & 1.08 & $-$2.78 \\
4518.332 & 22.1 & 1.08 & $-$2.56 \\
4571.971 & 22.1 & 1.57 & $-$0.31 \\
4583.409 & 22.1 & 1.16 & $-$2.84 \\
4609.265 & 22.1 & 1.18 & $-$3.32 \\
4657.201 & 22.1 & 1.24 & $-$2.29 \\
4708.663 & 22.1 & 1.24 & $-$2.35 \\
4764.525 & 22.1 & 1.24 & $-$2.69 \\
4798.531 & 22.1 & 1.08 & $-$2.66 \\
4865.610 & 22.1 & 1.11 & $-$2.70 \\
4874.009 & 22.1 & 3.09 & $-$0.86 \\ 
4911.194 & 22.1 & 3.12 & $-$0.64 \\
5069.090 & 22.1 & 3.12 & $-$1.62 \\
5185.902 & 22.1 & 1.89 & $-$1.41 \\
5211.530 & 22.1 & 2.59 & $-$1.41 \\
5336.786 & 22.1 & 1.58 & $-$1.60 \\
5381.022 & 22.1 & 1.56 & $-$1.97 \\
5396.247 & 22.1 & 1.58 & $-$3.18 \\
5418.768 & 22.1 & 1.58 & $-$2.13 \\
6680.134 & 22.1 & 3.09 & $-$1.89 \\ 
5238.960 & 24.0 & 2.71 & $-$1.43 \\
5304.180 & 24.0 & 3.46 & $-$0.77 \\
6330.090 & 24.0 & 0.94 & $-$2.90 \\
4848.230 & 24.1 & 3.86 & $-$1.13 \\
5237.330 & 24.1 & 4.07 & $-$1.18 \\
4007.270 & 26.0 & 2.76 & $-$1.66 \\
4010.180 & 26.0 & 3.64 & $-$2.03 \\
4014.270 & 26.0 & 3.02 & $-$2.33 \\
4080.880 & 26.0 & 3.65 & $-$1.54 \\
4423.840 & 26.0 & 3.65 & $-$1.61 \\
4547.850 & 26.0 & 3.55 & $-$1.01 \\
4587.130 & 26.0 & 3.57 & $-$1.74 \\
4602.000 & 26.0 & 1.61 & $-$3.15 \\
4630.120 & 26.0 & 2.28 & $-$2.59 \\
4635.850 & 26.0 & 2.85 & $-$2.36 \\ 
4690.140 & 26.0 & 3.69 & $-$1.64 \\
4704.950 & 26.0 & 3.69 & $-$1.57 \\
4733.590 & 26.0 & 1.49 & $-$2.99 \\
4745.800 & 26.0 & 3.65 & $-$1.27 \\
4779.440 & 26.0 & 3.42 & $-$2.02 \\
4787.830 & 26.0 & 3.00 & $-$2.60 \\
4788.760 & 26.0 & 3.24 & $-$1.76 \\
4799.410 & 26.0 & 3.64 & $-$2.23 \\
4802.880 & 26.0 & 3.64 & $-$1.51 \\
4807.710 & 26.0 & 3.37 & $-$2.15 \\
4808.150 & 26.0 & 3.25 & $-$2.79 \\
4809.940 & 26.0 & 3.57 & $-$2.72 \\
4835.870 & 26.0 & 4.10 & $-$1.50 \\
4839.540 & 26.0 & 3.27 & $-$1.82 \\
4844.010 & 26.0 & 3.55 & $-$2.05 \\
4875.880 & 26.0 & 3.33 & $-$2.02 \\
4882.140 & 26.0 & 3.42 & $-$1.64 \\
4892.860 & 26.0 & 4.22 & $-$1.29 \\
4907.730 & 26.0 & 3.43 & $-$1.84 \\
4918.010 & 26.0 & 4.23 & $-$1.36 \\
4946.390 & 26.0 & 3.37 & $-$1.17 \\
4950.100 & 26.0 & 3.42 & $-$1.49 \\
4994.130 & 26.0 & 0.92 & $-$3.06 \\
5198.710 & 26.0 & 2.22 & $-$2.13 \\
5225.530 & 26.0 & 0.11 & $-$4.79 \\
5247.050 & 26.0 & 0.09 & $-$4.95 \\
5250.210 & 26.0 & 0.12 & $-$4.93 \\
5295.310 & 26.0 & 4.42 & $-$1.59 \\
5373.710 & 26.0 & 4.47 & $-$0.71 \\
5379.570 & 26.0 & 3.69 & $-$1.51 \\
5386.330 & 26.0 & 4.15 & $-$1.67 \\
5441.340 & 26.0 & 4.31 & $-$1.63 \\
5466.400 & 26.0 & 4.37 & $-$0.63 \\
5466.990 & 26.0 & 3.57 & $-$2.23 \\
5491.830 & 26.0 & 4.19 & $-$2.19 \\
5554.890 & 26.0 & 4.55 & $-$0.27 \\
5560.210 & 26.0 & 4.43 & $-$1.09 \\
5618.630 & 26.0 & 4.21 & $-$1.25 \\
5638.260 & 26.0 & 4.22 & $-$0.72 \\
5651.470 & 26.0 & 4.47 & $-$1.90 \\
5679.020 & 26.0 & 4.65 & $-$0.82 \\
5705.460 & 26.0 & 4.30 & $-$1.35 \\
5731.760 & 26.0 & 4.26 & $-$1.20 \\
5852.220 & 26.0 & 4.55 & $-$1.23 \\
5855.080 & 26.0 & 4.61 & $-$1.48 \\
5956.690 & 26.0 & 0.86 & $-$4.60 \\
5987.070 & 26.0 & 4.80 & $-$0.43 \\
6005.540 & 26.0 & 2.59 & $-$3.60 \\
6065.480 & 26.0 & 2.61 & $-$1.53 \\
6079.010 & 26.0 & 4.65 & $-$1.02 \\
6082.710 & 26.0 & 2.22 & $-$3.58 \\
6093.640 & 26.0 & 4.61 & $-$1.40 \\
6096.670 & 26.0 & 3.98 & $-$1.83 \\
6151.620 & 26.0 & 2.18 & $-$3.29 \\
6165.360 & 26.0 & 4.14 & $-$1.47 \\
6173.340 & 26.0 & 2.22 & $-$2.88 \\
6187.990 & 26.0 & 3.94 & $-$1.62 \\
6200.310 & 26.0 & 2.61 & $-$2.43 \\
6213.430 & 26.0 & 2.22 & $-$2.48 \\
6219.280 & 26.0 & 2.20 & $-$2.43 \\
6226.740 & 26.0 & 3.88 & $-$2.12 \\
6232.640 & 26.0 & 3.65 & $-$1.24 \\
6380.740 & 26.0 & 4.19 & $-$1.38 \\
6430.850 & 26.0 & 2.18 & $-$2.00 \\
6593.870 & 26.0 & 2.43 & $-$2.42 \\
6597.560 & 26.0 & 4.80 & $-$0.97 \\
6625.020 & 26.0 & 1.01 & $-$5.34 \\
6703.570 & 26.0 & 2.76 & $-$3.06 \\
6705.100 & 26.0 & 4.61 & $-$0.87 \\
6710.320 & 26.0 & 1.49 & $-$4.76 \\
6713.750 & 26.0 & 4.80 & $-$1.50 \\
6725.360 & 26.0 & 4.10 & $-$2.10 \\
6726.670 & 26.0 & 4.61 & $-$1.13 \\
6739.520 & 26.0 & 1.56 & $-$4.79 \\
6750.150 & 26.0 & 2.42 & $-$2.62 \\
6793.260 & 26.0 & 4.08 & $-$2.33 \\
4508.290 & 26.1 & 2.86 & $-$2.35 \\
4576.340 & 26.1 & 2.84 & $-$2.98 \\
4582.830 & 26.1 & 2.84 & $-$3.22 \\
4620.520 & 26.1 & 2.83 & $-$3.31 \\
4629.340 & 26.1 & 2.81 & $-$2.48 \\
4635.320 & 26.1 & 5.96 & $-$1.58 \\
4670.180 & 26.1 & 2.58 & $-$4.07 \\
4993.350 & 26.1 & 2.81 & $-$3.68 \\
5234.620 & 26.1 & 3.22 & $-$2.18 \\
5264.800 & 26.1 & 3.23 & $-$3.13 \\
5414.070 & 26.1 & 3.22 & $-$3.58 \\
6084.090 & 26.1 & 3.20 & $-$3.88 \\
6149.240 & 26.1 & 3.89 & $-$2.84 \\
6247.550 & 26.1 & 3.89 & $-$2.44 \\
6369.460 & 26.1 & 2.89 & $-$4.11 \\
6432.680 & 26.1 & 2.89 & $-$3.57 \\
6456.380 & 26.1 & 3.90 & $-$2.19 \\
4904.410 & 28.0 & 3.54 & $-$0.25 \\
4953.200 & 28.0 & 3.74 & $-$0.68 \\
4998.220 & 28.0 & 3.61 & $-$0.79 \\
5084.090 & 28.0 & 3.68 & $-$0.07 \\
5088.530 & 28.0 & 3.85 & $-$1.06 \\
5115.390 & 28.0 & 3.83 & $-$0.13 \\
5593.730 & 28.0 & 3.90 & $-$0.77 \\
5748.350 & 28.0 & 1.68 & $-$3.24 \\
5846.990 & 28.0 & 1.68 & $-$3.45 \\
5996.730 & 28.0 & 4.24 & $-$1.06 \\ 
6086.280 & 28.0 & 4.27 & $-$0.45 \\
6111.070 & 28.0 & 4.09 & $-$0.83 \\
6130.130 & 28.0 & 4.27 & $-$0.89 \\
6204.600 & 28.0 & 4.09 & $-$1.15 \\
6223.980 & 28.0 & 4.11 & $-$0.97 \\
6322.160 & 28.0 & 4.15 & $-$1.21 \\
4810.528 & 30.0 & 4.08 & $-$0.16 \\
\end{longtable}
\end{center}

\clearpage
\twocolumn

\end{appendix}


\end{document}